\documentclass[aps,superscriptaddress,floats,floatfix,eqsecnum,nofootinbib]{revtex4}
\usepackage{epsfig,graphicx,amssymb,amsbsy,bm,amsmath,rotating}
\usepackage{hyperref}
\usepackage{psfrag}

\newcommand{\vx}{\ensuremath{\vec{x}}}
\newcommand{\vk}{\ensuremath{\vec{k}}}

\newcommand{\be}{\begin{equation}}
\newcommand{\ee}{\end{equation}}
\newcommand{\bea}{\begin{eqnarray}}
\newcommand{\eea}{\end{eqnarray}}
\begin{document}
\title{CMB quadrupole suppression: I. Initial conditions of inflationary perturbations. }
\author{D. Boyanovsky}
\email{boyan@pitt.edu} \affiliation{Department of Physics and
Astronomy, University of Pittsburgh, Pittsburgh, Pennsylvania 15260,
USA} \affiliation{Observatoire de Paris, LERMA. Laboratoire
Associ\'e au CNRS UMR 8112.
 \\61, Avenue de l'Observatoire, 75014 Paris, France.}
\affiliation{LPTHE, Universit\'e Pierre et Marie Curie (Paris VI) et
Denis Diderot (Paris VII), Laboratoire Associ\'e au CNRS UMR 7589,
Tour 24, 5\`eme. \'etage, 4, Place Jussieu, 75252 Paris, Cedex 05,
France}
\author{H. J. de Vega}
\email{devega@lpthe.jussieu.fr} \affiliation{LPTHE, Universit\'e
Pierre et Marie Curie (Paris VI) et Denis Diderot (Paris VII),
Laboratoire Associ\'e au CNRS UMR 7589, Tour 24, 5\`eme. \'etage, 4,
Place Jussieu, 75252 Paris, Cedex 05,
France}\affiliation{Observatoire de Paris, LERMA. Laboratoire
Associ\'e au CNRS UMR 8112.
 \\61, Avenue de l'Observatoire, 75014 Paris, France.}
\affiliation{Department of Physics and Astronomy, University of
Pittsburgh, Pittsburgh, Pennsylvania 15260, USA}
\author{N. G. Sanchez}
\email{Norma.Sanchez@obspm.fr} \affiliation{Observatoire de Paris,
LERMA. Laboratoire Associ\'e au CNRS UMR 8112.
 \\61, Avenue de l'Observatoire, 75014 Paris, France.}
\date{\today}
\begin{abstract}
We investigate the issue of initial conditions of   curvature and
tensor perturbations at the beginning of slow roll inflation and
their effect on the power spectra. Renormalizability and small back
reaction constrain the high $ k $ behavior of the Bogoliubov
coefficients that define these initial conditions.  We introduce a
\emph{transfer function } $ D(k) $ which encodes the effect of
generic initial conditions on the power spectra. The constraint from
renormalizability and small back reaction entails that $ D(k)
\lesssim \mu^2/k^2 $ for large $ k $, implying that observable
effects from initial conditions are more prominent in the \emph{low}
multipoles. This behavior affects  the CMB quadrupole by the
observed amount $ \sim 10-20\% $ when $ \mu $ is of the order of the
energy scale of inflation. The effects on high $l$-multipoles are
suppressed by a factor $ \sim 1/l^2 $ due to the fall off  of $ D(k) $
for large wavevectors $ k $. We show that the determination of
generic initial conditions for the fluctuations is equivalent to the
scattering problem by a potential $\mathcal{V}(\eta)$ localized just
prior to the slow roll stage. Such potential leads to a transfer
function $ D(k) $ which automatically obeys the renormalizability and
small backreaction constraints.  We find that an attractive
potential $ \mathcal{V}(\eta) $ yields a {\bf suppression} of the
lower CMB multipoles. Both for curvature and tensor modes, the
quadrupole suppression depends only on the energy scale of $
\mathcal{V}(\eta) $,  and on the time interval where $ \mathcal
{V}(\eta) $ is nonzero. A suppression of the quadrupole for
curvature perturbations consistent with the data is obtained when
the scale of the potential is of the order of $k^2_Q$ where $k_Q$ is
the wavevector whose physical wavelength is the Hubble radius today.
\end{abstract}


\maketitle
\tableofcontents
\section{Introduction}

 Inflation is a central part of early Universe cosmology
originally introduced to explain several shortcomings of the
standard Big Bang cosmology \cite{guth}-\cite{riottorev} and at the
same time it provides a mechanism for generating scalar (density)
and tensor (gravitational wave) perturbations
\cite{mukha}-\cite{CMBgiova}. A distinct aspect of inflationary
perturbations is that metric perturbations are generated by quantum
fluctuations of the scalar field(s) that drive inflation. After
their wavelength becomes larger than the Hubble radius, these
fluctuations are amplified and grow, becoming classical and
decoupling from  causal microphysical processes. Upon re-entering
the horizon, during the matter era, these classical perturbations
seed the inhomogeneities which generate structure upon gravitational
collapse \cite{mukha}-\cite{CMBgiova}.

\medskip

Most inflationary models
predict fairly generic features: a gaussian, nearly scale invariant
spectrum of (mostly) adiabatic scalar and tensor primordial
fluctuations, making the inflationary paradigm fairly robust. The
confirmation of many of the predictions of inflation by current
high precision observations places inflationary cosmology on solid
grounds.

\medskip

The gaussian, adiabatic and nearly scale invariant spectrum
of primordial fluctuations provide an excellent fit to the highly
precise wealth of data provided by the Wilkinson Microwave
Anisotropy Probe (WMAP)\cite{bennett}-\cite{WMAP33}.
Perhaps the most striking validation of inflation as a mechanism for
generating \emph{superhorizon} (`acausal')  fluctuations is the
anticorrelation peak in the temperature-polarization (TE) angular
power spectrum at $ l \sim 150 $ corresponding to superhorizon
scales \cite{kogut,peiris}.

\medskip

The power spectra for scalar curvature and tensor (gravitational
wave) quantum fluctuations generated during the inflationary stage
determine the angular power spectrum of CMB anisotropies. Their
initial conditions are usually chosen as Bunch-Davies conditions
\cite{BD}, which  fix the asymptotic behavior for large negative
conformal time $\eta$ to be the same as in Minkowski space time in
term of positive frequencies. The Bunch-Davies states transform as
irreps under the maximal symmetry group $ O(4,1) $ of de Sitter
space-time. (Other initial states were also considered
\cite{diferen}). The requirement that the energy momentum tensor be
renormalizable constrains the UV asymptotic behaviour of the
Bogoliubov coefficients that encode different  initial conditions
\cite{motto2}.

\medskip

The possibility that more precise observations of the anisotropies
in the cosmic microwave background (CMB) may probe physical aspects
of the initial conditions of quantum fluctuations motivated a
substantial effort to study different initial conditions and their
potential observational consequences \cite{otros}.
The remarkable high quality data and the exhaustive analysis of the
three year results from WMAP \cite{WMAP31} reveal  that outlaying
points and wiggles near the acoustic peaks in earlier data have all but disappeared
thus rendering much less statistical significance to potential
observables from `transplanckian' effects in the
CMB \cite{otros} on small angular scales.

\medskip

On the other hand, while there are no statistically significant departures
from the slow roll inflationary scenario at small angular scales
($ l\gtrsim 100 $), the third year WMAP data again confirms the
surprinsingly low quadrupole $ C_2 $ \cite{WMAP31}-\cite{WMAP33} and
suggests  that it cannot be completely explained by galactic
foreground contamination.
The low value of the quadrupole has been
an intriguing feature on large angular scales since first observed
by COBE/DMR \cite{cobe}, and confirmed by the WMAP
data \cite{bennett}-\cite{WMAP33}.
The observation of a low
quadrupole \cite{tegmark} and the surprising alignment of quadrupole
and octupole \cite{tegmark,virgo} sparked many different proposals
for their explanation \cite{expla}. More recently the robustness of
these features in the low multipoles to foreground contamination
has been studied \cite{tegmark2} with the suggestion \cite{abramoc2}
that these may originate in extended (large scale) foregrounds
perhaps generated by SZ distortions by hot electrons in the local
supercluster.

\bigskip

\textbf{The goals and main results  :}

\bigskip

In this article we address the issue of the initial conditions of
the fluctuations and the effects they imprint on the primordial
spectra of curvature and tensor perturbations within the
\emph{effective} field theory of inflation \cite{1sN}. We show that
initial conditions consistent with renormalizability and small back
reaction influence mainly the low CMB multipoles. In particular we
find a \emph{suppression} of CMB quadrupole  consistent with the
current observations. Furthermore, we formulate the problem of
the initial conditions in terms of a potential that affects the
evolution of scalar and tensor fluctuations \emph{prior to slow
roll}. In a companion article\cite{II} we show that this potential
is a generic feature of a brief \emph{fast roll} stage   prior to
slow roll inflation, and study its observational consequences as a
suppression in the CMB quadrupole for temperature and tensor modes.
We highlight that these results are derived within the context of
the effective theory of inflation\cite{1sN,clar}. Namely, we provide
a consistent assessment of the initial conditions at  the energy
scale of inflation which is the grand unification scale ($ \sim 10^{16}$GeV), 
without the {\bf need} to advocate
transplanckian physics or extra assumptions as an explanation for
non-BD initial conditions. As described in detail below, non-BD
initial conditions can be consistently incorporated within the
effective field theory valid at the inflation scale.

\medskip

The goal of this article is to study the effects on the power
spectra of curvature and gravitational wave perturbations of initial
conditions consistent with the criteria of renormalizability of the
gauge invariant energy momentum tensor and negligible back reaction.
These general initial conditions are related to the Bunch-Davies
initial conditions by a Bogoliubov transformation. The
renormalizability criteria constrains the high $ k $ behavior of the
Bogoliubov coefficients\cite{motto2}. We show that these constraints
imply that observable effects from initial conditions  are more
pronounced in the \emph{low multipoles}, namely in the region of the
angular power spectra corresponding to the Sachs-Wolfe plateau.  Our
main results are summarized as follows:

\begin{itemize}
\item{We introduce a \emph{transfer function for initial conditions} $ D(k) $
which encodes the effect of general initial conditions on the power
spectra. The constraint from renormalizability and small back
reaction entail that $ D(k) = {\cal O}(\mu^2/k^2) $ for large $ k $. We
show that this behavior naturally yields an observable correction to
the quadrupole. This correction can account for the 
{\bf suppression} of the quadrupole by the observed amount $ \sim 10-20\% $
when the high energy tail of the initial conditions is set by the
inflation scale. The corrections to higher $ l $ multipoles are
suppresed by a factor $ \sim 1/l^2 $ and therefore they are not
observable within the present data.}

\item{The equation for the fluctuations can be interpreted
as a one dimensional Schr\"odinger equation with a (conformal) time
dependent potential. We argue that this potential features two
distinct parts: (i) the slow roll part $ [(\nu^2 -1/4 )/ \eta^2 ] $
which is repulsive, (like a repulsive potential barrier, $ \eta $
being the conformal time, $ \nu $ being $ 3/2 $ plus slow roll
parameters), and (ii) a different part $ \mathcal V(\eta) $ with
support before slow roll starts. The potential $ \mathcal V(\eta) $
vanishes in the slow roll stage, hence it does not affect  the
dynamics during this stage, but its presence imprints the physical
initial conditions to the fluctuations in the slow roll stage both
for metric and tensor perturbations.}

\item{  We demonstrate that the problem of setting generic initial
conditions in the fluctuations equation is equivalent to the
scattering problem by a potential. Thus, by implementing the
powerful methods of scattering theory we show that the potential $
\mathcal V(\eta) $  yields initial conditions on the fluctuations
for the beginning of slow roll whose large $ k $ behavior is
consistent with renormalizability. We describe the potential
$\mathcal V(\eta)$ in a general manner and establish that an
\emph{attractive} potential $\mathcal V(\eta)$ leads to an
observable \emph{suppression} of the quadrupole. }

\item{We find that the effects  on the power spectrum are robust and only  depend on
the strength and width of  the potential $ \mathcal V(\eta) $,
namely   on the energy scale of $ \mathcal V(\eta) $, which is the
inflation scale, and on the time interval where $ \mathcal V(\eta) $
is nonzero.}

\item{Our analysis applies both to the curvature as well as the tensor
fluctuations. Therefore, we predict that the initial conditions for
slow roll also affect the quadrupole for B-modes.}

\end{itemize}

We show in the companion article \cite{II} that the potential $
\mathcal V(\eta) $ is quite generic and originates in a stage of
\emph {fast roll} inflaton dynamics. This is an early stage during
which the inflaton varies rapidly, slowing down to merge with the
slow roll stage.

\section{Initial conditions and the energy momentum tensor of
scalar and tensor perturbations.}\label{sec:inicon}

The effective field theory of slow roll inflation has two main
ingredients: the classical Friedmann equations in terms of a
\emph{classical} part of the energy momentum tensor described
by a homogeneous and isotropic condensate,  and a quantum part.
The latter   features   scalar fluctuations determined by a gauge
invariant combination of the scalar field (inflaton)
and metric fluctuations, and a tensor component,
gravitational waves. A consistency condition for this description  
is that the contributions from the fluctuations to the
energy momentum tensor be much smaller than those from the
homogeneous and isotropic condensate. The effective field theory
must include renormalization counterterms so that it is insensitive
to the possible ultraviolet singularities of the short wavelength
fluctuations. Different initial conditions on the mode functions of
the quantum fluctuations yield different values for their
contribution to the energy momentum tensor.
Different initial conditions on the mode functions of the quantum
fluctuations yield different values for the energy momentum tensor.

Criteria  for acceptable initial conditions must include the
following: i) back reaction effects from the quantum fluctuations
should not modify the inflationary dynamics described by the
inflaton, ii) the ultraviolet counterterms that renormalize  the
energy momentum tensor should not depend on the particular choice of
initial conditions, namely different initial conditions \emph{should
not} introduce new ultraviolet divergences: a single renormalization
scheme, independent of initial conditions, should render the energy
momentum tensor UV finite. This set of criteria imply that the
ultraviolet allowed states have their large $ k $ behaviour
constrained up to the fourth order in $ 1/k $ \cite{motto2}. In ref.
\cite{motto2} only the energy momentum tensor of \emph{inflaton}
fluctuations was considered. However, the fluctuations of the
inflaton field are \emph{not} gauge invariant, and in order to
establish a set of criteria for UV allowed initial states in a gauge
invariant manner, we must study the full gauge invariant energy
momentum of scalar and tensor fluctuations.

\subsection{Scalar perturbations}\label{sec:scalar}

The gauge invariant energy momentum tensor for quadratic scalar
metric fluctuations has been obtained in ref.\cite{abramo} where the
reader is referred to for details. Its form simplifies in
longitudinal gauge, and in cosmic time it is given  by\cite{abramo}
\bea
\langle T_{00} \rangle = && M^2_{Pl} \Bigg[12 \; H  \; \langle
\psi \dot{\psi} \rangle - 3 \; \langle (\dot{\psi})^2 \rangle +
\frac{9}{a^2(t)} \; \langle (\nabla \psi)^2\rangle \Bigg] +\nonumber \\
& & + \frac12 \; \langle (\dot{ \phi})^2\rangle + \frac{\langle
(\nabla  \phi)^2\rangle}{2\,a^2(t)} + \frac{V''(\Phi )}2 \; \langle
\phi^2 \rangle + 2 \; V'(\Phi ) \; \langle \psi \, \phi \rangle
\label{Too} \; , \eea \noindent where $ \Phi (t) $ stands for the
zero mode of the inflaton field, $ \phi(t,{\vec x}) $ for the
inflaton fluctuations around $ \Phi (t) , \; \psi(t,{\vec x}) $ is
the longitudinal gauge form of the Bardeen potential and the dots
stand for derivatives with respect to cosmic time. During inflation
the Newtonian potential and the Bardeen potential are the same in
the longitudinal gauge\cite{mukhanov,riotto} and this property has
been used in the above expression.

In longitudinal gauge, the equations of motion in cosmic time for
the Fourier modes are\cite{mukhanov,riotto} \bea \label{phieq}
&&\ddot{\psi}_{\vec k}+ \left(H -2 \;
\frac{\ddot{\Phi}}{\dot{\Phi} }\right)\dot{\psi}_{\vec k}+ \left[2
\; \left(\dot{H } -2 \; H \; \frac{\ddot{\Phi} }{\dot{\Phi}
}\right)+ \frac{k^2}{a^2(t)}\right]{\psi}_{\vec k}=0  \; , \cr \cr
&& \label{delfieqn} \ddot{ \phi}_{\vec k}+3 \; H \; \dot{
\phi}_{\vec k}+\left[V''[\Phi ]+\frac{k^2}{a^2(t)} \right]
\phi_{\vec k}+2 \; V'[\Phi ] \; \psi_{\vec k}- 4 \; \dot{\Phi}  \;
\dot{\psi}_{\vec k}=0  \; \; , \eea \noindent with the constraint
equation \be \label{vin} \dot{\psi}_{\vec k}+H \; \psi_{\vec k}=
\frac1{2M^2_{Pl}} \;  \phi_{\vec k} \; \dot{\Phi}   \; . \ee Initial
conditions on the mode functions of the quantum fluctuations
correspond to an initial value problem at a fixed time hypersurface.
For modes  of cosmological relevance this time slice at which the
initial conditions are established is such that these modes are
\emph{subhorizon}.  Therefore, we must focus on the contribution to
the energy momentum tensor from subhorizon fluctuations, and in
particular in the large momentum region to assess the criteria for
UV allowed states.

For subhorizon modes with wavevectors $ k \gg a(t) \;  H $, the
solutions of the equation (\ref{phieq}) are \cite{mukhanov} 
\be
\label{apro} \psi_{\vec k}(t) \approx e^{\pm i k \eta} \Rightarrow
\dot{\psi}_{\vec k}(t) \sim \frac{i \, k}{a(t)} \; {\psi}_{\vec
k}(t) \; . 
\ee 
For $ k \gg a(t) \; H $ the constraint equation
(\ref{vin}) entails that \cite{abramo} 
\be \label{conss} 
\psi_{\vec
k}(t) \approx \frac{i \, a(t)}{2\,M^2_{Pl}\,k} \; \dot{\Phi}   \;
\phi_{\vec k} \; . \ee In slow roll, \be \label{fidotSR} \dot{\Phi}
= - \frac{V'(\Phi )}{3 \; H } \left[ 1 + {\cal
O}\left(\frac1{N}\right) \right] = - H \; M_{Pl} \;  \sqrt{2 \;
\epsilon_v} \left[ 1 + {\cal O}\left(\frac1{N}\right)\right] \; ,
\ee where the slow roll parameters $ \epsilon_v, \; \eta_v $ are of
the order $ 1/N $ \cite{1sN}, \be \epsilon_v = \frac{M^2_{Pl}}2 \;
\left[\frac{V'(\Phi )}{V(\Phi )} \right]^2  = {\cal
O}\left(\frac1{N}\right)  \quad , \quad
\eta_v = M^2_{Pl} \; \frac{V''(\Phi )}{V(\Phi )}  = {\cal
O}\left(\frac1{N}\right) \; , \label{slowroll} 
\ee 
and $ N \sim 55 $
stands for the number of efolds from horizon exit until the end of
inflation. Therefore, for subhorizon modes, \be \psi_{\vec k}(t)
\approx -i \; \sqrt{2 \; \epsilon_v} \; \left(\frac{H \;  a(t) }{k}
\right) \frac{\phi_{\vk}}{2 \; M_{Pl}} \; . \ee These identities,
valid in the limit $ k \gg a(t) \; H $ allow to obtain an estimate
for the different contributions to $ T_{00} $. The first line of
eq.(\ref{Too}), namely the contribution from the Newtonian potential
mode with comoving wavevector $ k $ is \be \langle
T^{(\psi)}_{00}\rangle \approx 6 \, \epsilon_v \, H^2 \langle (
\phi_{\vk} )^2 \rangle \label{Tphi} \; . \ee The first three terms
in the second line of eq.(\ref{Too}) (the quadratic contribution
from the scalar field fluctuations) is \be \label{Tvarphi} \langle
T^{\phi}_{00} \rangle \approx \left(\frac{k}{a(t)}\right)^2 \langle
( \phi_{\vk})^2 \rangle \; , \ee and the crossed term is: \be
\label{Tphivarphi} V'(\Phi) \;  \langle \psi_k   \; \phi_k \rangle
\approx \epsilon_v \, H^2 \,  \left(\frac{a(t) \;
H}{k}\right)\langle ( \phi_{\vk})^2 \rangle \; , \ee Therefore, in
slow roll, $ \epsilon_v, \; \eta_v \ll 1 $ and for subhorizon modes
$ k \gtrsim a(t) H $, the leading contribution to the energy
momentum tensor for the scalar fluctuations is given by the
contribution from the inflaton fluctuations, namely \be \langle
T_{00} \rangle   \simeq \frac12 \; \langle (\dot{ \phi})^2\rangle +
\frac{\langle (\nabla \phi)^2\rangle}{2\,a^2(t)} + \frac{V''(\Phi
)}2 \; \langle \phi^2\rangle \,. \label{Toolead} \ee Furthermore, in
terms of the slow roll parameter $ \eta_v , \; V''= 3 \; \eta_v \;
H^2 $ and for subhorizon wavevectors with $ k\gg a(t) \; H $ the
last term in eq.(\ref{Toolead}) is subdominant and will be
neglected. Hence, the contribution to the energy momentum tensor
\emph{from subhorizon fluctuations during the slow roll stage} is
determined by the subhorizon quantum fluctuations of the inflaton
and given by \be \label{T00sub} \langle T_{00} \rangle   \simeq
\frac12 \; \langle (\dot{ \phi})^2\rangle + \frac{\langle (\nabla
\phi)^2\rangle}{2\,a^2(t)} \; . \ee This analysis allows us to
connect with the  the results in ref.\cite{motto2} for inflaton
fluctuations.

The  inflaton fluctuation obeys the equation of motion \be
\label{eqnofmot} \ddot{ \phi}_{\vk} +3 \; H \;
\dot{\phi}_{\vk}+\left[3 \,H^2 \,\eta_v +\frac{k^2}{a^2(t)} \right]
\phi_{\vk}= 0 \; . \ee In what follows it is convenient to pass to
conformal time in terms of which, the homogeneous and isotropic FRW
metric is determined by \be ds^2 = dt^2-a^2(t)(d\vec x)^2 =
C^2(\eta)[d\eta^2 - (d\vec x)^2] \; , \ee where $t$ and $\eta$ stand
for cosmic and conformal time respectively. During slow roll 
\be
C(\eta) = -\frac1{\eta H  \; (1-\epsilon_v)}  \label{CSR} \; . 
\ee
In conformal time $ \eta $ the solution of eq.(\ref{eqnofmot}) is
given by 
\be \phi_{\vec{k}}(\eta) =
\frac1{C(\eta)}\left[\alpha_{\vk} \; S_\phi(k,\eta)+
\alpha^{\dagger}_{-\vk} \; S_\phi^*(k,\eta)\right] \; ,
\label{expalfa} 
\ee 
where the mode functions $ S_\phi(k,\eta) $ are solutions of 
\be \left[\frac{d^2}{d\eta^2}+k^2 +M^2_{\Phi} \; C^2(\eta)-
\frac{C''(\eta)}{C(\eta)} \right]S_\phi(k,\eta) = 0  \label{phieqn}
\; , \ee here, \be \label{mass2} M^2_{\Phi} = V''(\Phi ) = 3 \;  H^2 \;
\eta_v \; . 
\ee 
and prime stands for derivative with respect to the
conformal time. Using eqs.(\ref{slowroll}) and (\ref{CSR}) during
slow roll, this equation simplifies to 
\be
\left[\frac{d^2}{d\eta^2}+k^2 - \frac{\nu^2_\phi -\frac14}{\eta^2}
\right]S_\phi(k,\eta) = 0  \; , \label{fluqeq} \ee where for
inflaton fluctuations during slow-roll \be \label{Seqn2} \nu_\phi =
\frac32+\epsilon_v -\eta_v +  {\cal O}\left(\frac1{N^2}\right)\; .
\ee 
The operators $ \alpha_{\vk}, \; \alpha^{\dagger}_{\vk} $ in eq.
(\ref{expalfa}) obey the usual canonical commutation relations.

\subsection{Tensor perturbations}\label{sec:tensor}

Tensor perturbations (gravitational waves) are gauge invariant.
The expectation value
of their  energy momentum pseudo-tensor in a quantum state
  has been obtained in ref.\cite{abramo} (see also ref.\cite{giova2})
  and is given by
\be \label{tensorT00}
\langle T^{(T)}_{00} \rangle = M^2_{Pl} \Bigg\{ H  \; \langle
  \dot{h}_{kl} h_{kl} \rangle + \frac18
\left[ \langle \dot{h}_{kl} \; \dot{h}_{kl}\rangle + \frac1{a^2(t)}\langle
  \nabla h_{kl} \;  \nabla h_{kl}  \rangle \right] \Bigg\} \; ,
\ee where the dot stands for derivative with respect to cosmic time.
The equations of motion for the spatial Fourier transform of the
dimensionless tensor field $ h $ are \be \label{heqn}
\ddot{h}_{kl}(\vk)+ 3\;  H\;  \dot{h}_{kl}(\vk) + \frac{k^2}{a^2(t)}
\;  h_{kl}(\vk) = 0\, . \ee Tensor perturbations correspond to
minimally coupled  massless fields with two physical polarizations.
Passing to conformal time the spatial Fourier transform of the
quantum fields are written as \cite{rmp,CMBgiova} \be
h_{ij}(\vk,\eta) = \frac1{C(\eta) \; M_{Pl} \;  \sqrt2 }
\sum_{\lambda=\times,+} \epsilon_{ij}(\lambda,\vec{k})
\left[\alpha_{\lambda,\vec{k}} \; \, S_T(k;\eta)+
\alpha^\dagger_{\lambda,\vec{k}} \; \, S^*_T(k;\eta) \right] \; ,
\label{tens} \ee \noindent where $ \lambda $ labels the two standard
transverse and traceless polarizations $ \times $ and $ + $. The
operators $ \alpha_{\lambda,\vec{k}}, \;
\alpha^\dagger_{\lambda,\vec{k}} $ obey the usual canonical
commutation relations, and $ \epsilon_{ij}(\lambda,\vec{k}) $ are
the two independent traceless-transverse tensors constructed from
the two independent polarization vectors transverse to $
\hat{\bf{k}} $, chosen to be real and normalized such that $
\epsilon^i_j(\lambda,\vec{k})\, \;
\epsilon^j_k(\lambda',\vec{k})=\delta^i_k
 \; \delta_{\lambda,\lambda'} $.

\medskip

The mode functions $ S_T(k;\eta) $ obey the differential equation
\be\label{Sten} S^{''}_{T}(k;\eta)+\left[k^2-
\frac{C''(\eta)}{C(\eta)}\right]S_{T}(k;\eta) = 0 \; , \ee where in
the slow roll regime, \be \label{eqnC}\frac{C''(\eta)}{C(\eta)} =
\frac{\nu^2_{T}-\frac14}{\eta^2}~~;~~\nu_T = \frac32+\epsilon_v +
{\cal O}\left(\frac1{N^2}\right)\; . \ee Thus, to leading order in
slow roll the mode functions for gravitational waves obey the same
equations of motion as for scalar fields but with vanishing mass,
namely setting $ \eta_v =0 $.

\subsection{Initial conditions}

We treat both scalar and tensor perturbations on the same footing by
focusing on mode functions solutions of the general equation \be
\left[\frac{d^2}{d\eta^2}+k^2 - \frac{\nu^2 -\frac14}{\eta^2}
\right]S(k,\eta) = 0 \; . \label{geneq} \ee For general initial
conditions we write \be \label{genS} S (k;\eta) = A (k)\,g_{\nu
}(k;\eta) + B (k) \,f_{\nu }(k;\eta) \; . \ee where two linearly
independent solutions of eq.(\ref{geneq}) are, \bea g_{\nu }(k;\eta)
& = & \frac12 \; i^{\nu +\frac12} \;
\sqrt{-\pi \eta}\,H^{(1)}_{\nu }(-k\eta) \; , \label{gnu}\\
f_{\nu }(k;\eta) & = & [g_{\nu }(k;\eta)]^*\label{fnu}  \; ,
\eea
\noindent where $ H^{(1)}_\nu(z) $ are Hankel functions.  These
solutions are normalized so that their Wronskian is given by
\be
W[g_\nu(k;\eta),f_\nu(k;\eta)]=
g'_\nu(k;\eta) \; f_\nu(k;\eta)-g_\nu(k;\eta) \; f'_\nu(k;\eta) = -i \; .
\label{wronskian}
\ee
(here prime stands for derivative with respect to the conformal time).
For the specific cases of scalar or tensor perturbations, the mode
functions  and coefficients $ A(k), \; B(k) $ will feature a subscript index
$ \phi, \; T $, respectively.

For wavevectors deep inside the Hubble radius $ | k \, \eta | \gg 1
$, the mode functions for arbitrary $ \nu $ have the Bunch-Davies
asymptotic behavior
\be
g_{\nu}(k;\eta) \buildrel{\eta \to
-\infty}\over= \frac1{\sqrt{2 \, k}} \; e^{-ik\eta} \quad ,  \quad
f_{\nu}(k;\eta) \buildrel{\eta \to -\infty}\over=
\frac1{\sqrt{2k}} \; e^{ ik\eta} \; , \label{fnuasy}
\ee
and for $ \eta \to 0^- $, the mode functions behave as:
\be\label{geta0}
g_\nu(k;\eta)\buildrel{\eta \to 0^-}\over=
\frac{\Gamma(\nu)}{\sqrt{2 \, \pi \; k}} \; \left(\frac2{i \; k \;
\eta} \right)^{\nu - \frac12} \; .
\ee
The complex conjugate formula holds for $ f_{\nu}(k;\eta) $.

In particular, in the scale invariant case $ \nu=\frac32 $ which is
the leading order in the slow roll expansion, the mode functions eqs.(\ref{gnu})
simplify to
\be
g_{\frac32}(k;\eta) =
\frac{e^{-ik\eta}}{\sqrt{2k}}\left[1-\frac{i}{k\eta}\right]\,.\label{g32}
\ee
The coefficients $ A(k), \; {B}(k) $  for the general solution
eq.(\ref{genS}) are determined by an initial condition on the mode
functions $ S(k;\eta) $ at a given initial conformal time
$ {\bar \eta} $, namely
\bea
B (k)& = & -i[g_{\nu }(k;{\bar \eta}) \; S' (k;{\bar \eta})-g'_{\nu
}(k;{\bar \eta}) \; S (k;{\bar \eta})]
\label{B} \\
A (k )& = & -i[f'_{\nu }(k;{\bar \eta}) \; S (k;{\bar \eta})-f_{\nu
}(k;{\bar \eta}) \; S' (k;{\bar \eta})] \label{A} \; . \eea
Canonical commutation relations for the Heisenberg fields entail
that \be |A (k)|^2-| B (k)|^2 = 1 \label{constraint} \; . \ee The
S-vacuum state $ |0\rangle_\mathcal{S} $ is annihilated by the
operators $ \alpha_{\vk} $ associated with the modes $ S (k;\eta) $.
A different choice of the coefficients $ A (k); \; B(k) $ determines
different choices of vacua, the Bunch-Davies vacuum corresponds to $
A(k)=1, \;  B(k)=0 $.   An  illuminating representation of these
coefficients can be gleaned by computing the expectation value of
the number operator in the Bunch-Davies vacuum. Consider the
expansion of the fluctuation field both in terms of Bunch-Davies
modes $ g_\nu(k;\eta) $ and in terms of the general modes $ S
(k;\eta) $, for example for the scalar field $ \phi $ (similarly for
tensor fields with a subscript $ T $ and corresponding
normalization)
\begin{equation}\label{expa}
\phi_{\vec{k}}(\eta) =
\frac1{C(\eta)}\left[ a_{\vk} \;  g_{\nu_\phi}(k;\eta) +
a^{\dagger}_{-\vk} \; g^*_{\nu_\phi}(k;\eta)\right] \equiv
\frac1{C(\eta)}\left[\alpha_{\vk} \; S_\phi(k;\eta)+
\alpha^\dagger_{-\vk} \; S^*_\phi(k;\eta)\right] \; ,
\end{equation}
\noindent the creation and annihilation operators are related by a
Bogoliubov transformation
\be
\alpha^\dagger_{\vk}  =  A_\phi(k) \;  a^\dagger_{\vk} -
B_\phi(k) \;  a_{-\vk} \quad , \quad
\alpha_{\vk}  =  { A}^*_\phi(k) \;  a_{\vk} -  {B}^*_\phi(k) \;
a^\dagger_{-\vk}  \; .
\ee
The Bunch-Davis vacuum $ |0\rangle_{BD} $ is annihilated by $ a_{\vk} $,
hence we find the expectation value
\be
{}_{BD}\langle 0|\alpha^\dagger_{\vk} \; \alpha_{\vk}
|0\rangle_{BD}= | B_\phi(k)|^2= N_\phi(k)  \; .
\ee
Where $ N_\phi(k) $ is interpreted as the number of S-vacuum particles
in the Bunch-Davis vacuum. In combination with the constraint
eq.(\ref{constraint}) the above result suggests the following
representation for the  coefficients $ A(k); \; B(k) $
\be
A_\phi(k) = \sqrt{1+N_\phi(k)} \; e^{i\theta_A(k)}~~;~~
  B_\phi(k)=\sqrt{N_\phi(k)} \; e^{i\theta_{B}(k)} \label{bogonum} \; ,
\ee \noindent where $ N_\phi(k), \; \theta_{A,{B}}(k) $ are real.
The only relevant phase is the difference \be \label{phasediff}
\theta_k=\theta_{B}(k)-\theta_A(k) \; . \ee Notice that we provide
the initial conditions at a given conformal time $ {\bar \eta} $
which is obviously the same for all $k$-modes. This is the
consistent manner to define the initial value problem (or Cauchy
problem) for the fluctuations. This is different from what is often
done in the literature when an ad-hoc dependence on $ k $ is given
to $ {\bar \eta} $ \cite{otros}.

\subsection{The Transfer Function of Initial Conditions and its Asymptotic Behaviour}

For gauge invariant scalar perturbations, the analysis leading to
eq.(\ref{T00sub}) indicates that in order to study
the energy momentum tensor for general initial conditions
it is enough to
consider the leading order in the slow roll expansion.
Consistently with neglecting the contributions from the Newtonian
potential as well as the term proportional to $ V''[\Phi] $ for the inflaton
fluctuations, we set $ \nu=3/2 $ in the expression for the mode
functions eq.(\ref{gnu}).
This simplification results in  considering the scalar
field fluctuations as massless and minimally coupled to gravity.

The energy density in the vacuum state defined by the new initial
conditions is
\be
\rho = {}_\mathcal{S}\langle 0| T_{00}|0 \rangle_\mathcal{S} \label{rho}\,.
\ee
The renormalized energy
density from the fluctuations of the inflaton field is found to
be\cite{motto2}
\be
\rho = \rho^{BD}+ I_1 + I_2 \label{rhoren}
\ee
where $\rho^{BD}$ corresponds to the Bunch-Davies vacuum initial conditions
$ N_k=0 $ and
\bea 
I_1  & = & \frac1{2 \, \pi^2} \int_0^\infty dk\,k^2\,
\left\{N_\phi(k)\, |\dot{F}(k,\eta)|^2+
\sqrt{N_\phi(k)[1+N_\phi(k)]}\,
\mathrm{Re}\left[e^{-i\theta_k}\left(\dot{F}(k,\eta)\right)^2\right]
\right\} \label{I1}\\
I_2  & = & \frac1{2 \, \pi^2} \int_0^\infty dk\,k^2\,\frac{k^2}{a^2}
\left\{N_\phi(k) \,| {F}(k,\eta)|^2+ \sqrt{N_\phi(k)[1+N_\phi(k)]}\,
\mathrm{Re}\left[e^{-i\theta_k}\left( {F}(k,\eta)\right)^2\right]
\right\} \label{I2}
\eea
where $ F(k,\eta) $ is given in terms of the Bunch-Davis mode function
eq.(\ref{g32}) for $ \nu=3/2 $ as
\be
F(k,\eta) =  (-H\,\eta) \; g_{\frac32}(k,\eta)   =
\frac{H}{\sqrt{2 \; k^3}} \; e^{-ik\eta} \; (i-k \; \eta) \label{BDmode} \; \,.
\ee
The power spectrum of the inflaton fluctuations is given by \cite{liddle,rmp},
\be
P_{\phi}(k,t) = {}_\mathcal{S}\langle0||\phi_k(\eta)|^2|0\rangle_\mathcal{S} =
P^{BD}_{\phi}(k,t)+ \frac{k^3}{\pi^2}\left\{ N_\phi(k) \; |
{F}(k,\eta)|^2+ \sqrt{N_\phi(k)[1+N_\phi(k)]} \;
\mathrm{Re}\left[e^{-i\theta_k}\left({F}(k,\eta)\right)^2\right]\right\} \; ,
\ee
where we used eq.(\ref{expa}) and
\be
P^{BD}_{\phi}(k,t) = \frac{k^3}{2\pi^2} \; | {F}(k,\eta)|^2 \; .
\ee
We find,
\bea \label{I1fin}
&&I_1= \frac{(H\eta)^4}{(2\pi)^2} \int_0^\infty dk\, k^3 \left\{
N_\phi(k)-\sqrt{N_\phi(k)[1+N_\phi(k)]}\cos[2 \, k \, \eta+\theta_k]
\right\} \\ \cr
&& I_2 =\frac{(H^2\eta)^2}{(2\pi)^2} \int_0^\infty dk \, k  \left\{
N_\phi(k)\,(1+k^2 \, \eta^2)-\sqrt{N_\phi(k)[1+N_\phi(k)]}\left[(1-k^2 \, \eta^2)
\cos[2 \, k \, \eta+\theta_k]+2k\eta
\sin[2 \, k \, \eta+\theta_k]\right]\right\}\label{I2fin}  \; , \\ \cr
&& P_{\phi}(k,t)= \left(\frac{H}{2\pi}\right)^2
\left\{(1+k^2 \, \eta^2)[1+2 \; N_\phi(k)]-2 \; \sqrt{N_\phi(k)[1+N_\phi(k)]}
\left[(1-k^2 \, \eta^2)\cos[2 \, k \, \eta+\theta_k]+2 \, k \, \eta
\sin[2 \, k \, \eta+\theta_k]\right] \right\} \nonumber  \; .
\eea
Evaluating the power spectrum after horizon crossing $ |k \, \eta| \ll
1 $, yields
\be
\frac{P_\phi}{P^{BD}_\phi}\Bigg|_{|k\eta| \ll 1} = 1+
D_\phi(k)\label{Powerratio}
\ee
where we have introduced the \emph{transfer function for initial conditions}
\be
D_\phi(k)= 2 \; | {B}_\phi(k)|^2 -2 \;
\mathrm{Re}\left[A_\phi(k)B^*_\phi(k)\right] =
2 \; N_\phi(k)-2 \; \sqrt{N_\phi(k)[1+N_\phi(k)]} \;  \cos \theta_k
 \label{Dofk}  \; .
\ee 
The integrals $ I_{1,2} $ are finite provided that
asymptotically for $ k\rightarrow \infty $ the occupation numbers
behave as \be \label{adiabatic}
N_\phi(k)=\mathcal{O}\left(\frac1{k^{4+\delta}}\right) \; , \ee with
$ \delta >0 $. Namely, the finiteness of the energy momentum tensor
constrains the asymptotic behaviour of the occupation numbers to
vanish faster than $ \frac1{k^4} $ for $ k \to \infty $
\cite{motto2}. Of course, this asymptotic condition leaves a large
freedom on the occupation numbers $ N_k $.

We systematically impose the constraint eq.(\ref{adiabatic}) which guarantees
the finiteness of energy momentum tensor. This is not always the
case for initial conditions considered in the literature (see \cite{otros}).

Let us establish a bound on the large momentum behavior of $ N_k $
inserting the asymptotic behavior \be N_k = N_\mu
\left(\frac{\mu}{k}\right)^{4+\delta}  \; , \label{ocupa} \ee with $
0< \delta \ll 1 $ in the integrals $ I_{1,2} $. Namely, assuming
that the integrals are dominated by the region of high momenta $ k/H
\gg 1 $ and that the occupation number attains the largest possible
values consistent with ultraviolet finiteness
[eq.(\ref{adiabatic})]. We observe that  $ k \; |\eta| \gg 1 $ in
the early stages of inflation for large $ k $, and that the maximum
contribution from these integrals are at early time $ \eta \sim -1/H
$. Hence, the oscillatory terms in $ I_1, \; I_2 $ average out and
we have from eqs.(\ref{I1fin})-(\ref{I2fin}), \be \label{integ} I_1
\sim I_2 \sim \frac{N_\mu}{(2\pi)^2} \,\frac{\mu^4}{\delta}  \; .
\ee The contribution from the fluctuations to the energy momentum
tensor does not lead to large back reaction effects affecting the
inflationary dynamics provided that $ I_1, \; I_2 \ll  H^2 \;
M^2_{Pl} $, which yields \be  \label{Nmu} {N_\mu} \ll 2 \, \pi^2 \;
\frac{H^2 \; M^2_{Pl}}{\mu^4} \; \delta \; . \ee

Eq.(\ref{Nmu}) provides an occupation number distribution exhibiting
the largest asymptotic value compatible with an UV finite energy
momentum tensor. This maximal occupation number distribution falls
off for $ k \to \infty $ with the minimal acceptable power tail
exponent $ 4 + \delta $ with $ \delta \ll 1 $.

\medskip

Gravitons are massless particles with two independent polarizations,
therefore their energy momentum tensor is given by eq.(\ref{T00sub})
times a factor two. The first term in the energy momentum
pseudotensor for gravitational waves eq.(\ref{tensorT00}) features only
one time derivative, which results in only one factor $ k $ for large
momenta, whereas the terms with two time or spatial derivatives
yield $ k^2 $. Therefore, the first term is subdominant in the
ultraviolet and the short wavelength  contribution to the energy
momentum (pseudo) tensor of gravitational waves is the same as that
for a free massless scalar field, up to a factor 2 from the physical
polarization states \cite{giova2,nuestroveff}. Therefore we can
directly extend the results obtained above for scalar fluctuations
to the case of tensor fluctuations.

\medskip

Small backreaction effects from the fluctuations is a necessary
consistency condition for the validity of the inflationary scenario.
In addition, the condition that different initial states
\emph{should not affect} the renormalization aspects of the energy
momentum tensor is a consistency condition on the renormalizability
of the effective field theory of inflation: the theory should be
insensitive to the short distance physics for \emph{any} initial
conditions. These criteria lead to the following important
consequences:

\begin{itemize}
\item{If $ \mu \sim M_{Pl} $ then $ N_\mu  \lesssim H^2/M^2_{Pl}   \ll 1 $
because $ H/M_{Pl} \ll 1 $  in the effective field theory expansion
and the effect of initial conditions becomes negligible. }

\item{For $ \mu \sim \sqrt{H\,M_{Pl}} \sim
V^{\frac14}(\Phi) $, namely  $ \mu $ of the order of the scale of
inflation during the slow roll stage, then $ N_{\mu} \lesssim 1 $.
For example for $\delta \sim 0.01$ one obtains   $ N_\mu \sim 0.1 $.
If $ \mu \ll \sqrt{H\,M_{Pl}} $, for example $ \mu \sim H $, the
bound eq.(\ref{Nmu}) is rather loose allowing a wide range of $
N_\mu $ with potentially appreciable effects. }

\item{The condition that the occupation number falls off faster
than $ 1/k^4 $ for large wavevector, implies that
the possible effects from different initial conditions are more
prominent for the smaller wavevectors, those that exited the Hubble radius
the \emph{earliest}. For cosmologically relevant wavevectors, these
are those that crossed about $55$ e-folds before the end of inflation.
Today those wavevectors correspond to the present Hubble scale,
hence the low multipoles in the CMB.  }
\end{itemize}

We conclude that consistent with renormalizability and small back
reaction there may be a substantial effect from the initial
conditions when the characteristic scale $ \mu \leq \sqrt{H\;
M_{Pl}} $. The rapid fall-off of the occupation numbers $ N_\phi(k)
$ for subhorizon wavelengths and the back-reaction constraint
eq.(\ref{Nmu}) entails that for these modes the transfer function
eq.(\ref{Dofk}) for initial conditions simplifies to \be
\label{simpleD} D_\phi(k) \buildrel{N_\phi(k) \ll 1}\over=
 -2 \; \sqrt{N_\phi(k)} \; \cos\theta_k \; ,
\ee
and that the \emph{smaller values} of $ k $
yield the \emph{larger corrections  from initial conditions}. The
result eq.(\ref{simpleD}) suggests a \emph{suppression} of the power
spectrum for $ \cos\theta_k >0 $.  These observations will be crucial
below when we study the effect of initial conditions on the
multipoles of the CMB.

While this discussion focused on the
fluctuations of the inflaton, they are directly applicable to the
case of gauge invariant perturbations studied below.

\section{Effects of the Initial conditions on the Curvature Perturbations}

In the previous section we focused on the backreaction effects from
initial conditions beginning with the gauge invariant energy
momentum tensor for scalar and tensor perturbations. Since the
fluctuation modes are initialized on a fixed time hypersurface while
their wavelength are well inside the Hubble radius, we established a
correspondence with ref.\cite{motto2} which refer solely to the
quantum fluctuations of the inflaton field. The effect of different
initial conditions is encoded in the Bogoliubov coefficients, and in
particular in the occupation numbers $ N_k $ and the phases $
\theta_k $. Ultraviolet allowed initial conditions require that $
N_k $ diminishes faster than $ 1/k^4 $ for asymptotically large
momenta. Small backreaction effects require in general that $ N_k
\ll 1$.

In this section we study UV allowed initial conditions on the
quantum fluctuations associated with gauge invariant variables which
determine the power spectra of observables. We focus on both  scalar
and tensor perturbations.

\subsection{ Effects of Initial Conditions on the Power Spectrum}\label{curper}

The gauge invariant curvature perturbation of the comoving
hypersurfaces is given in terms of the Newtonian potential ($\psi$)
 and inflaton fluctuation  ($\phi$) by\cite{mukhanov,rmp}
\be \label{curvature} \mathcal{R}= -\psi-\frac{H}{\dot{\Phi} } \;
\phi  \; . \ee where $ \dot \Phi  $ stands for the derivative of the
inflaton field $ \Phi  $ with respect to the cosmic time $ t $.

It is convenient to introduce the gauge invariant potential
\cite{mukhanov}, \be u(\vx,t)=-z \;  \mathcal{R}(\vx,t) \label{u}
\; , \ee where \be \label{za} z= a(t) \; \frac{\dot{\Phi} }{H} \;  .
\ee The spatial Fourier transform of the gauge invariant field $
u(\vx,t) $ is expanded in terms of conformal time mode functions and
creation and annihilation operators as follows\cite{mukhanov,rmp}
\be \label{curvau} u(\vk,\eta)=
\alpha_\mathcal{R}(k)\,S_\mathcal{R}(k;\eta)+
\alpha^\dagger_\mathcal{R}(k)\,S^*_\mathcal{R}(k;\eta) 
\ee 
where the
vacuum state is annihilated by the operators $ \alpha_\mathcal{R}(k)
$ and the mode functions are solutions of the equation 
\be
\Bigg[\frac{d^2}{d\eta^2}+k^2- \frac1{z} \frac{d^2z}{d\eta^2}
\Bigg]S_\mathcal{R}(k;\eta) =0 \,. \label{Scureq} 
\ee 
During slow roll and to leading order in slow roll variables 
\be \label{eqnz}
\frac1{z} \frac{d^2z}{d\eta^2} = \frac2{\eta^2} \Bigg[1+ \frac32\; (3
\; \epsilon_v-\eta_v) \Bigg] = \frac{\nu^2_\mathcal{R}-
\frac14}{\eta^2} \quad , \quad \nu_\mathcal{R} = \frac32+ 3 \,
\epsilon_v -\eta_v +  {\cal O}\left(\frac1{N^2}\right) \; . 
\ee 
The general solution of eq.(\ref{Scureq}) in the slow roll regime is
given by \be S_\mathcal{R}(k;\eta) = A_\mathcal{R}(k) \;
g_{\nu_\mathcal{R}}(\eta) + B_\mathcal{R}(k) \;
g^*_{\nu_\mathcal{R}}(\eta) \; , 
\ee 
where the function $
g_\nu(\eta) $ is given by eq.(\ref{gnu}), the Bogoliubov
coefficients obey the relation (\ref{constraint}) and can be written
in terms of the occupation number of Bunch-Davis particles as in
eq.(\ref{bogonum}) with the label $ \mathcal{R} $ replacing $ \phi $.

The power spectrum of curvature perturbations in the state with general
initial conditions is given by \cite{liddle,rmp},
\be\label{curvapot}
P_\mathcal{R}(k) \buildrel{\eta \to  0^-}\over=
\frac{k^3}{2 \; \pi^2} \; 
 \Big|\frac{S_\mathcal{R}(k;\eta)}{z} \Big|^2 \; .
\ee 
From eq.(\ref{eqnz}) we see that in the slow-roll regime $ z $
behaves  as \be\label{zeta2} z(\eta) = \frac{z_0}{ (-k_0 \;
\eta)^{\nu_R-\frac12}} \; , \ee where $ z_0 $ is the value of $ z $
when a reference scale $ k_0 $ exits the horizon. Combining this
result with the small $\eta$ limit eq.(\ref{geta0}) we find from
eqs.(\ref{curvapot}) and (\ref{zeta2}), \be P_{\mathcal{R}}(k) =
P^{BD}_{\mathcal{R}}(k)\Big[1+ D_\mathcal{R}(k) \Big] \label{powR}
\ee where we introduced the transfer function for the initial
conditions of curvature perturbations: 
\bea \label{DofkR}
D_\mathcal{R}(k) &=& 2 \; | {B}_\mathcal{R}(k)|^2 -2 \;
\mathrm{Re}\left[A_\mathcal{R}(k) \;
B^*_\mathcal{R}(k)\,i^{2\nu_R-3}\right] = \cr \cr &=& 2 \;
N_\mathcal{R}(k)-2 \; \sqrt{N_\mathcal{R}(k)[1+N_\mathcal{R}(k)]} \;
\cos\left[ \theta_k - \pi \left(\nu_R - \frac32 \right) \right] 
\eea
and 
\be\label{potinBD} 
P^{BD}_\mathcal{R}(k) = \left( \frac{k}{2 \,
k_0}\right)^{3 - 2 \; \nu_\mathcal{R}} \;
\frac{\Gamma^2(\nu_\mathcal{R})}{\pi^3} \; \left(\frac{k \; H}{a(t)
\; {\dot \Phi}}\right)_0^2  \; . 
\ee 
The index $ 0 $ refers to the
time where the reference scale $ k_0 $ exits the horizon. In terms
of the  slow-roll parameter $ \epsilon_v $   this expression
simplifies to the usual result\cite{liddle,rmp} 
\be  \label{potBD}
P^{BD}_\mathcal{R}(k) = \left( \frac{k}{2 \, k_0}\right)^{n_s - 1}
\; \frac{\Gamma^2(\nu_\mathcal{R})}{\pi^3} \; \frac{H^2}{2 \;
\epsilon_v  \; M_{Pl}^2 }\equiv \mathcal{A}^2_\mathcal{R} \;
\left(\frac{k}{k_0}\right)^{n_s - 1} \; , 
\ee 
where the amplitude is given by 
\be \label{ampR} 
\mathcal{A}^2_\mathcal{R} = \frac1{8 \,
\pi^2 \; \epsilon_v } \; \left(\frac{H}{M_{Pl}}\right)^2 \left\{ 1 +
(3 \; \epsilon_v - \eta_v  )\left[\ln 4 +
\psi\left(\frac32\right)\right]+
 {\cal O}\left(\frac1{N^2}\right)\right\} \; ,
\ee
$ n_s - 1 = 3 - 2 \; \nu_\mathcal{R} = 2 \; \eta_v  - 6 \;
\epsilon_v  $ in the slow-roll regime,  $ \psi(z) $ is the digamma
function and $ \psi\left(\frac32\right) = - 1.463510026.....\;$ . As
shown above, for wavevectors of cosmological relevance,
$ N_\mathcal{R}(k)\ll 1 $, hence to lowest order in slow roll
($ 2 \; \nu_\mathcal{R}=3 $), the transfer function for initial conditions
simplifies to
\be \label{Dcurvat}
D_\mathcal{R}(k) \buildrel{N_\mathcal{R}(k)\ll 1}\over=
-2 \; \sqrt{N_\mathcal{R}(k)}\,\cos\theta_k \; .
\ee
Therefore, for a positive $ \cos \theta_k $, we have a {\bf negative}
$ D_\mathcal{R}(k) $. That is, the initial conditions eq.(\ref{DofkR})
{\bf suppress} the power in such case.

\subsection{Tensor perturbations (Gravitational waves)}

The quantization of tensor fluctuations for general initial
conditions has been discussed in section \ref{sec:tensor}.

Following the same steps as in sec. \ref{curper} we find the power
spectrum of gravitational waves to be \cite{liddle,rmp},
\be \label{powh}
P_T(k)  \buildrel{\eta \to 0^-}\over= \frac{k^3}{2
\; \pi^2} \;  \Big|\frac{S_T(k;\eta)}{C(\eta)} \Big|^2 
=  P^{BD}_{T}(k)\Big[1+ D_T(k) \Big] \; ,
\ee
where the transfer function for the initial conditions of tensor
perturbations is
\be  \label{Dofkh}
D_T(k) = 2 \; | {B}_T(k)|^2 -2 \;
\mathrm{Re}\left[A_T(k)B^*_T(k)\,i^{2\nu_T-3}\right] =
2 \; N_T(k)-2 \;\sqrt{N_T(k)[1+N_T(k)]}\,
\cos\left[ \theta_k - \pi (\nu_T - \frac32 ) \right]
\ee and \be \label{powh2} P^{BD}_{T}(k) = \mathcal{A}^2_T \;
\left(\frac{k}{k_0}\right)^{n_T} \; , \ee with \be n_T= -2 \;
\epsilon_v \quad , \quad
\frac{\mathcal{A}^2_T}{\mathcal{A}^2_\mathcal{R}} = r = 16 \;
\epsilon_v \; . \ee The contribution from gravitational waves to the
energy momentum (pseudo) tensor is gauge invariant and up to a
factor of two from the polarizations is exactly of the form
eq.(\ref{T00sub}) with $ \phi $ replaced by $ h $
\cite{nuestroveff}. Thus, the constraint on the occupation number
eq.(\ref{ocupa})-(\ref{Nmu}) from the analysis of the backreaction
and renormalizability translate \emph{directly} to the
case of gravitational waves for the occupation number $ N_T(k) $. \\
This implies
that corrections to the power spectrum
of tensor modes from initial conditions are substantial if $\mu$, the  
asymptotic $k$ scale of $ N_T(k) $, is
$ \mu \leq  \sqrt{H \; M_{Pl}} $, that is  of the order of the inflation scale
[see discussion below eq.(\ref{ocupa})]. \\
We get from eq.(\ref{Dofkh})
for $ N_T(k) \ll 1 $ and to leading order in slow roll,
\be \label{Dh2}
D_T(k) \buildrel{N_T(k)\ll 1}\over= -2 \; \sqrt{N_T(k)} \; \cos\theta_k \; .
\ee
Again, for a positive $ \cos \theta_k $, we have a {\bf negative}
$ D_T(k) $. That is, the initial conditions {\bf suppress} the tensor power
spectrum in such case.

\section{The effect of initial conditions on the low multipoles of the CMB}

We have shown above that the fast fall off of the occupation number
$ N(k) $ (for the corresponding perturbation) entails that initial
conditions can only provide substantial corrections for perturbation
modes whose wavevectors crossed out of the Hubble radius
\emph{early} during inflation. In turn, today  these wavevectors
correspond to scales of the order of the  Hubble radius, namely to
the low multipoles in the CMB.

In the region of the Sachs-Wolfe plateau for $ l \lesssim 30 $, the
matter-radiation transfer function can be set equal to unity and the
$ C_l's $ are given by\cite{mukhanov}-\cite{CMBgiova} \be C_l
=\frac{4\pi}{9} \int_0^\infty \frac{dk}{k}\,  {P}_X(k) \left\{
j_l[k(\eta_{tod}-\eta_{LSS})]\right\}^2 \label{Cls} \; , \ee where $
P_X $ is the power spectrum of the corresponding perturbation, $
X=\mathcal{R} $ for curvature perturbations and $ X=T $ for tensor
perturbations,  $ j_l(x) $ are spherical Bessel functions
\cite{abra} and \be \eta_{tod}-\eta_{LSS} = \frac1{a_0 \;
H_0}\,\int^1_{\frac1{1+z_{LSS}}}
\frac{dx}{x\,\left[\frac{\Omega_{M}}{x}+
\Omega_{\Lambda}x^2\right]^{\frac12}} \; , \ee is the comoving
distance between today and the last scattering surface (LSS). In the
above expression we consider that the dark energy component is
described by a cosmological constant. Taking $ \Omega_M =0.28, \;
\Omega_{\Lambda}=0.72, \;  z_{LSS} =1100 $ we find \be
\eta_{tod}-\eta_{LSS} \sim \frac{3.3}{a_0 \;  H_0}  \; . \ee Notice
that $ k/[a_0 \; H_0] \sim d_H/\lambda_{phys}(t_0) $ is the ratio
between today's Hubble radius and the physical wavelength. The power
spectra for curvature ($\mathcal R$) or gravitational wave (T)
perturbations are of the form given by eqs.(\ref{powR}),
(\ref{potBD}), (\ref{powh}) and (\ref{powh2}), \be {P}_X =
\mathcal{A}^2_X \left( \frac{k}{k_0}\right)^{n_s-1}
\left[1+D_X(k)\right]\label{power} \; , \ee with $ n_s=n_\mathcal{R}
$ for curvature perturbations, $ n_s = 1+n_T $ for tensor
perturbations, and $ k_0 \sim a_0  \; H_0 $ is a pivot scale. Then,
from eq.(\ref{Powerratio}), the relative change $ \Delta C_l $ in
the $ C_l's $ due to the effect of generic initial conditions
(generic vacua), is given by \be C_l \equiv C ^{BD}_l + \Delta C_l
\quad , \quad \frac{\Delta C_l}{C_l} = \frac{\int^\infty_0
D_X(\kappa\,x)~ f_l(x)\,dx}{\int^\infty_0 f_l(x)\,dx} \label{DelC}
\ee where $ x= k/\kappa $ and \be \kappa \equiv a_0 \;  H_0/3.3 \;,
\label{kappa} \ee $ D(\kappa\,x) $ is the transfer function of
initial conditions for the corresponding perturbation and \be f_l(x)
= x^{n_s-2}[j_l(x)]^2 \,. \label{fls} \ee \noindent We now focus on
curvature perturbations since these are directly related to the
temperature fluctuations\cite{peiris}. For $ n_s =0.96 $
\cite{WMAP31}, the functions $ f_l(x) $ are strongly peaked at $ x
\sim l $. Therefore, $ \frac{\Delta C_l}{C_l} $ is dominated by
wavenumbers $ k \sim l \; \kappa $.

Low multipoles $ l $ correspond to  wavelengths \emph{today} of the
order of the Hubble radius. These wavelengths crossed the Hubble
radius about $ 55 $ e-folds before the end of inflation. Therefore,
since inflation lasted a total number of e-folds $ N_{total} \gtrsim
60 $, these wavevectors were subhorizon during the first few
e-folds, namely during the slow roll stage $ k\gg H $. As already
discussed, let us take for these wavevectors the occupation number $
N_k \ll 1 $  as given by the asymptotic expression \be N_k = N_\mu\,
\left(\frac{\mu}{k}\right)^{4+\delta}~~;~~0 < \delta \ll 1
\label{asynumber} \ee \noindent and assume that the angles $
\theta_k $ are slowly varying functions of $ k $ in the region of $
k $ corresponding to \emph{today's Hubble radius} so that
$\cos\theta_k \approx \overline{\cos \theta}$. Then, we find that
the fractional change in the coefficients $ C_l $ is given by \be
\frac{\Delta C_l}{C_l} \approx - 2 \; \sqrt{N_{\mu}} \;
\left(\frac{3.3 \; \mu}{a_0 \; H_0} \right)^{2+\frac{\delta}2} \;
\overline{\cos \theta}~~ \frac{I_l(n_s-2-\frac{\delta}2)}{I_l(n_s)}
\label{Claps} \ee where \be I_l(n_s) = \frac1{2^{3-n_s}} \;
\frac{\Gamma\left(3-n_s \right) \; \Gamma\left(\frac{2l+1-3+n_s+1}2
\right)}{\Gamma^2\left(\frac{4-n_s}2 \right) \;
\Gamma\left(\frac{2l+1+3-n_s+1}2 \right)} \label{Ilns}  \; . \ee To
obtain an estimate of the corrections, we take the values $ n_s=1,
\; \delta=0 $ and find \be \frac{\Delta C_l}{C_l} \approx - \frac43
\; \sqrt{N_{\mu}} \;  \left(\frac{3.3\,\mu}{a_0 \; H_0} \right)^ 2
\, \frac{\overline{\cos\theta}}{(l-1)(l+2)} \; . \label{Cl10} \ee
The $ \sim 1/l^2 $ behavior is a result of the $ 1/k^2 $ fall off of
$ D(k) $, a consequence of the renormalizability condition on the
occupation number. For the quadrupole, the relevant wave-vectors
correspond to $ x \sim 2 $, namely  $ k_Q \sim a_0 \; H_0 $. It is
convenient to write \be  \label{kH} k_Q =  a_{sr} \; H_{i} = a_0 \;
H_0 \; , \ee where $ a_{sr} $ and $ H_{i} $ are the scale factor and
the Hubble parameter during the slow roll stage of inflation when
the wavelength corresponding to \emph{today}'s Hubble radius exits
the horizon.

Hence, when the scale $ \mu $ in the asymptotic form of the
occupation number eq. (\ref {asynumber}) is of the order of the
largest scale of cosmological relevance \emph{today}, one has
\be
\frac{\mu}{a_0 \; H_0} \sim 1 \; ,
\ee
and for example with $ N_\mu \sim 0.01 $ we find that the fractional
change in the quadrupole is given by:
\be  \label{quad}
\frac{\Delta C_2}{C_2}\sim - 0.3 \; {\overline {\cos \theta}} \; .
\ee
namely a {\it suppression} of the order of $ \sim 10\% $
in the quadrupole provided that $ \overline{\cos\theta} \sim 1 $.
This corresponds to $ \mu $ of the order of the scale of
inflation during the slow roll stage [see eq.(\ref{kH})].

We emphasize that these are general arguments based on the criteria
of renormalizability and small backreaction which initial
conditions must fulfill.

In a companion article \cite{II} we show that these initial
conditions are effectively realized in inflationary dynamics. There
we show that a short stage just prior to the onset of slow roll
inflation and in which the inflaton field evolves \emph{fast},
imprints initial conditions on the curvature perturbations
corresponding to the above analysis.

\section{Initial Conditions as the Scattering by a potential} \label{condinfr}

In the previous sections we have systematically analyzed the consequences
of generic initial conditions different from Bunch-Davies, and which are UV
allowed and  consistent with small backreaction effects. We
now provide a novel explanation of the origin of these initial conditions.

The mode equations (\ref{geneq}) have the form of the Schr\"odinger
equation in one dimension suggesting to consider them more generally
as a \emph{potential scattering problem}.   The equations
(\ref{phieqn}), (\ref{Sten} ) and (\ref{geneq}) can be written in
the form, \be \label{sceq}
\Big[\frac{d^2}{d\eta^2}+k^2-W(\eta)\Big]S(k;\eta) =0 \ee as a
Schr\"odinger equation, with $ \eta $ the coordinate, $ k^2 $ the
energy and $ W(\eta) $ a potential that depends on the coordinate $
\eta $. In the cases under consideration \be W(\eta)  = \Bigg\{
\begin{array}{l}
z''/z  ~~\mathrm{for~curvature~perturbations}        \\ \\
C''/ C  ~~\mathrm{for~tensor~perturbations} \\
\end{array} \,.\label{defW}
\ee
It is convenient to  explicitly separate the behavior of $ W(\eta) $
during the slow roll stage by writing
\be\label{defV}
W(\eta)=  \mathcal{V}(\eta) +
\frac{\nu^2-\frac14}{\eta^2} \; ,
\ee
where $ \nu = \frac32+{\cal O}\left(\frac1{N}\right) $
[see eqs.(\ref{Seqn2}), (\ref{eqnC}) and (\ref{eqnz})]. Consider the potential
$ \mathcal{V}(\eta) $ localized in a region of conformal time
\emph{prior} to the  slow roll stage and which vanishes during the slow roll
inflationary stage (during which cosmologically relevant modes cross out of the
Hubble radius). Namely, a potential with the following properties:
\be \label{potloc}
\mathcal{V}(\eta) = \Bigg\{\begin{array}{l}
                \neq 0 ~\mathrm{for} ~-\infty < \eta < {\bar \eta} \\
                0 ~\mathrm{for} ~ {\bar \eta} < \eta   \\
              \end{array}
\ee where $ {\bar \eta} $ is the time when slow roll starts.

With Bunch-Davies (outgoing) boundary conditions,
$$
S(k;\eta) \buildrel{\eta \to -\infty}\over=  e^{-ik\eta}/\sqrt{2k}  \; ,
$$
the solution of eq.(\ref{sceq}) for $ W(\eta) $ given by eq.(\ref{defV}) and
eq.(\ref{potloc}) is
\be
S(k;\eta) = \Bigg\{ \begin{array}{l}
A(k) \; g_\nu(k)+B(k) \; g^*_\nu(k)  \; , ~~\mathrm{for}~  \eta >{\bar \eta} \\  \\
\frac{e^{-ik\eta}}{\sqrt{2k}} \; , ~~\mathrm{for}~ \eta \rightarrow -\infty  \\
\end{array} \label{Spo}
\ee
The coefficients $ A(k), \; B(k) $ are obtained by matching the wave
function $ S(k;\eta) $ and its derivative at $ {\bar \eta} $.  
$ A(k) $ and $ B(k) $
are simply related to the usual transmission and reflection coefficients of the
scattering by a potential (see below).

\medskip

We formally consider the conformal time starting at $ \eta = -\infty
$. However, the inflationary dynamics of the universe presumably
starts at some negative value $ \eta_0 < {\bar \eta} $.

In this article we study the general consequences of such potential,
deferring to a companion article\cite{II}   a comprehensive analytic
and numerical study that shows that an \emph{attractive} potential
of the form of eq.(\ref{potloc})  originates naturally within the
effective field theory of inflation from a brief \emph{fast roll
stage} just prior to slow roll.

\subsection{The effect of the potential  $ \mathcal{V}(\eta) $
as  a change in the initial conditions}

In summary, the equations for the quantum fluctuations are
\be
\left[\frac{d^2}{d\eta^2}+k^2-\frac{\nu^2-\frac14}{\eta^2}-
\mathcal{V}(\eta)  \right]S(k;\eta) = 0 \; .\label{eqnpsr2}
\ee
As discussed above the potential $ \mathcal{V}(\eta) $ describes the
deviation from the slow-roll dynamics during a (brief) stage prior
to slow roll and is vanishingly small for $ \eta > {\bar \eta} $, where $
{\bar \eta} $ denotes the beginning of the slow-roll stage during which
modes of cosmological relevance today exit the Hubble radius.

The retarded Green's function $ G_k(\eta,\eta') $ of
eq.(\ref{eqnpsr2}) for $ \mathcal{V}(\eta) \equiv 0 $   obeys
\be \label{green}
\left[\frac{d^2}{d\eta^2}+k^2-\frac{\nu^2-\frac14}{\eta^2}
 \right]G_k(\eta,\eta') = \delta(\eta-\eta') ~~;~~ G_k(\eta,\eta') =0~
\mathrm{for}~\eta'>\eta \quad ,
\ee
it  is given by
\be \label{Gret}
G_k(\eta,\eta') = i \left[g_\nu(k;\eta) \; g^*_\nu(k;\eta')-
g_\nu(k;\eta') \; g^*_\nu(k;\eta) \right] \Theta(\eta-\eta') \quad ,
\ee
where $ g_\nu(k;\eta) $ is given by eq.(\ref{gnu}).

The solution of the mode equation (\ref{eqnpsr2}) can be written as
an integral equation using the  Green's function
eq.(\ref{Gret}) 
\be \label{sol} 
S(k;\eta)= g_\nu(k;\eta) +
\int^{0}_{-\infty} G_k(\eta,\eta') \; \mathcal{V}(\eta') \;
S(k;\eta') \; d\eta' \;. 
\ee   
This is the   Lippmann-Schwinger
equation familiar in potential scattering theory. Inserting
eq.(\ref{Gret}) into eq.(\ref{sol}) yields, 
\be \label{solu}
S(k;\eta)= g_\nu(k;\eta) + i \; g_\nu(k;\eta)\,\int^{\eta}_{-\infty}
 g^*_\nu(k;\eta') \; \mathcal{V}(\eta') \;  S(k;\eta') \; d\eta'-i  \;
g^*_\nu(k;\eta)\,\int^{\eta}_{-\infty}
 g_\nu(k;\eta') \; \mathcal{V}(\eta') \;  S(k;\eta') \; d\eta'\quad .
\ee
 This solution has the Bunch-Davies asymptotic condition
\be S(k;\eta \rightarrow -\infty) = \frac{e^{-ik\eta}}{\sqrt{2k}}
\,. \ee Since $ \mathcal{V}(\eta) $ vanishes for  $ \eta > {\bar
\eta} $,  the mode functions $ S(k;\eta) $   for  $ \eta > {\bar
\eta} $ can be written as a linear combination  of the mode
functions $ g_\nu(k;\eta) $ and $ g^*_\nu(k;\eta) $ as follows, \be
\label{solSR} S(k;\eta) = A(k) \; g_\nu(k;\eta) + B(k) \;
g^*_\nu(k;\eta) \quad , \quad \eta > {\bar \eta} \quad , \ee where
the coefficients $ A(k) $ and $ B(k) $ can be read from
eq.(\ref{solu}), \bea A(k) & = &  1+ i\int^{0}_{-\infty}
g^*_\nu(k;\eta) \; \mathcal{V}(\eta) \; S(k;\eta) \;
d\eta\label{aofk} \cr \cr
 B(k) & = & -i \int^{0}_{-\infty}  g_\nu(k;\eta) \; \mathcal{V}(\eta)  \;
S(k;\eta) \;  d\eta\label{bofk}  \; .
\eea
The constancy of the Wronskian
$ W[g_\nu(\eta),g^*_\nu(\eta)]=-i $ and eq.(\ref{solSR}) imply the
relation
$$
|A(k)|^2-|B(k)|^2=1 \quad .
$$
It is clear  that the action of a potential $ \mathcal{V}(\eta) $
that vanishes for $ \eta > {\bar \eta} $ is equivalent to setting
initial conditions eqs.(\ref{solSR})-(\ref{bofk}) on the mode
functions at $ \eta = {\bar \eta} $   which subsequently evolve
during  the slow roll stage in  which $ \mathcal{V}(\eta) = 0 $.
This is one of our main observations.

\medskip

The integral equation can be solved iteratively in a perturbative
expansion if the potential $ \mathcal{V}(\eta) $
is small when compared to $ k^2 - (\nu^2-1/4)/\eta^2 $.
In such case, we can use for the coefficients $ A(k), \; B(k) $ the first
approximation obtained by replacing $ S (k;\eta') $ by
$ g_\nu(k;\eta') $ in the integrals eqs.(\ref{aofk})-(\ref{bofk}).
This is the Born approximation, in which
\be
A(k)   =    1+  i \int^{0}_{-\infty}
   \mathcal{V}(\eta)\,|g_\nu(k;\eta)|^2 \, d\eta   \quad  , \quad
 B(k)  =  - i  \int^{0}_{-\infty}
   \mathcal{V}(\eta)\, g^2_\nu(k;\eta)  \, d\eta \; .\label{bofk0}
\ee
These simple expressions are very illuminating. 
For asymptotically large $ k $
eq.(\ref{fnuasy}) for the mode functions can be used, and if the potential
$ \mathcal{V}(\eta) $ is differentiable and of compact support, an integration
by parts yields
 \be B(k) \buildrel{k \to \infty}\over=   -\frac1{4 \, k^2} \int^{0}_{-\infty}
 e^{-2 \, i \, k\eta} \; \mathcal{V}'(\eta) \, d\eta \label{bofk01} \; ,
\ee
where the prime stands for derivative with respect to $ \eta $. Therefore,
according to the Riemann-Lebesgue lemma,
$ N_k  = |B(k)|^2 \lesssim 1/k^4 $ for large $ k $ and UV
convergence in the integrals for the energy momentum tensor is guaranteed.
Hence, an immediate consequence of the explanation of
the initial conditions as a scattering problem with a localized
potential is that these initial conditions are \emph{automatically} 
ultraviolet allowed.

\medskip

To illustrate the main aspects and highlight the main consequences,
we consider now two simple potential models for $ \mathcal{V}(\eta)
$ localized at $ \eta = \eta_0 < {\bar \eta} $ and characterized by
a strength $ v_0 $ and width $ \Delta $. The first one   has an
exponential profile and the second is a square well. We solve the
first one in the Born approximation while the second one is exactly
solvable. These exact results agree remarkably well with the simpler
Born approximation. Thus, the \emph{exactly solvable} example
supports the reliability of the Born approximation in the present
framework.

\subsection{Born approximation}\label{sec:born}

As is clear from the integral equation (\ref{bofk}) the occupation
number $ N(k) $ grows with the strength of the potential $
\mathcal{V}(\eta) $. Moreover, as shown above, negligible
backreaction requires  $ N(k) \ll 1$  for wavevectors of
cosmological relevance. This is the regime where the Born
approximation eq.(\ref{bofk0}) is reliable.

To leading order in slow roll we consider the scale invariant case,
$ \nu=\frac32 $ with the mode functions given by eq.(\ref{g32}), and
model a potential localized at a time scale $ \eta_0 $ by \be
\mathcal{V}(\eta) = \frac{v_0}{\sqrt{\pi}}\, e^{- (\frac{\eta-
\eta_0}{\Delta})^2} \label{potential}  \; , \ee where
$\overline{\eta}=\eta_0+\Delta$. The localization length must be $
|\Delta| \ll |\eta_0| $ such that the potential must not influence
the dynamics during slow roll, and $ v_0 $ must be small for the
Born approximation to be valid. More precisely $ |v_0 \Delta| \ll
k$, as seen below. Under these conditions we find, \bea A(k) & = &
1+i\, \frac{v_0 \, |\Delta| }{2 \; k}\left[1+ \frac1{(k \,
\eta_0)^2}+
 \mathcal{O}\left(\Big|\frac{\Delta}{\eta_0}\Big|^2\right) \right] \; ,
\label{Aborn}\\
 B(k) & =  & -i \, \frac{v_0 \,  |\Delta| }{2 \, k}\,
 e^{-(k\Delta)^2} \, e^{- 2 \, i \, k \, \eta_0}
\left[1- \frac1{(k \, \eta_0)^2}-\frac{2 \, i}{(k \, \eta_0) } +
 \mathcal{O}\left(\Big|\frac{\Delta}{\eta_0}\Big|^2\right) \right] \; .
\label{Bborn}
\eea
To lowest order in $ v_0 $ the number of produced
Bunch-Davies (BD) modes $ N_k $ and the transfer function $ D(k) $ are given by
\be
N_k = \Big(\frac{v_0  \,|\Delta| }{2 \, k}\Big)^2 \, e^{-2 \,
(k \, \Delta)^2}\,\left[1+ \frac1{(k \, \eta_0)^2} \right]^2 \; ,
\label{Nborn}
\ee
and
\be
D(k) = -\frac{v_0 \;  |\Delta| }{ k} \,
e^{-(k \, \Delta)^2} \Bigg[\sin\big( 2 \, k \,
 |\eta_0|\big)\left(1-\frac1{(k \, \eta_0)^2} \right)+ \frac{2\cos\big( 2 \, k
  \, |\eta_0|\big)}{k \, |\eta_0|} \Bigg]\label{Dborn}  \; ,
\ee
respectively. The particle number $ N_k $ clearly falls off faster than
any power at large  $ k $, thereby ensuring the ultraviolet
convergence of the integrals in the energy momentum tensor.

This example reveals that a potential that is localized near  a
(conformal) time scale $ \eta_0$ results in a \emph{phase difference}
$ \sim e^{- 2 \, i \, k \, \eta_0} $ between the Bogoliubov
coefficients  $ A(k), \, B(k) $. This is a general result that stems
directly from the general expressions for these coefficients
eq.(\ref{bofk0}) and that in turn results in the oscillatory
component of the transfer function $ D(k) $.

In the integral eq.(\ref{DelC}) that yields the coefficients $
\Delta C_l/C_l $, the transfer function $ D(k) $ multiplies a
function that is strongly peaked at $ x \sim l $, namely, for
momenta $ k \sim l \; \kappa $. Therefore, if  $ k \; |\eta_0| \sim
l \; \kappa \;  |\eta_0| \gg 1 $, the rapid oscillations in $ D(k) $
average out in the integrand, resulting in a vanishing contribution
to the $ \Delta C_l/C_l's $. Hence, there are significant
contributions to $ \Delta C_l/C_l $ only  when $ l \; \kappa \;
|\eta_0| \sim 1 $. For the quadrupole this corresponds to, $ [a_0 \
H_0 \, |\eta_0|] \sim 1 $.

\medskip

The potential $ {\mathcal V}(\eta) $ acts prior to the slow roll
stage during which cosmologically relevant modes cross the Hubble
radius. For the corrections to the low multipoles be substantial,
the condition for wavevectors corresponding to the Hubble radius
today is $ k \sim  a_0 \, H_0  $ and $ k \,  |\eta_0| \sim 1 $. The
conclusion is that modifications to the low multipoles arise from a
potential $ \mathcal{V}(\eta)$ localized \emph{just prior to horizon
crossing of the modes whose wavelengths correspond to the Hubble
radius today}. It is also clear that the corrections for higher
wavevectors are strongly suppressed because of the rapid fall off of
$ B(k) $.

\begin{figure}[h]
\begin{center}
\includegraphics[height=3in,width=3in,keepaspectratio=true]{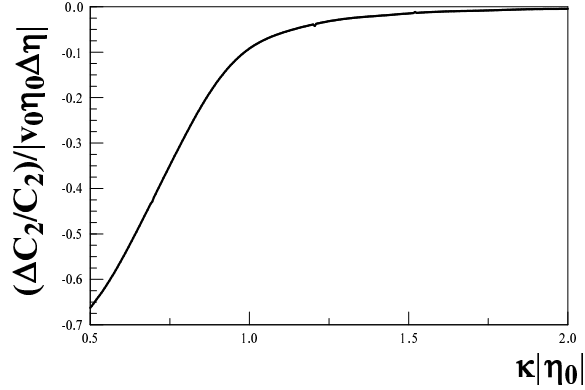}
\caption{The quadrupole correction $ (\Delta C_2/C_2)/|v_0
\eta_0\Delta| $ vs, $ \kappa |\eta_0|$  in the Born approximation
for an attractive potential of the form (\ref{potential}),  for $
|\Delta/\eta_0|=0.01, \; 0.05, \; 0.1 $. Here,    $ \kappa =
\frac{a_0 H_0}{3.3} $. It clearly reveals a suppression for $\kappa
\eta_0\sim 1$, that is for the modes whose wavelengths correspond to
the Hubble radius today} \label{fig:c2born}
\end{center}
\end{figure}

Furthermore, in the Born approximation the sign of the correction $
\Delta C_l/C_l $ is determined by the sign of the potential. In the
example given above with the potential eq.(\ref{potential}), it is
given by the sign of $v_0$, negative (positive) $ v_0 $
corresponding to an attractive (repulsive) potential.  Fig.
\ref{fig:c2born} shows the quadrupole correction $ \Delta C_2/C_2 $ determined
by the transfer function eq.(\ref{Dborn}) for an {\bf attractive}
potential ($ v_0=-|v_0| $) clearly revealing a {\bf suppression} for
$ \kappa|\eta_0| \sim 1 $.

\begin{figure}[h]
\begin{center}
\includegraphics[height=3in,width=3in,keepaspectratio=true]{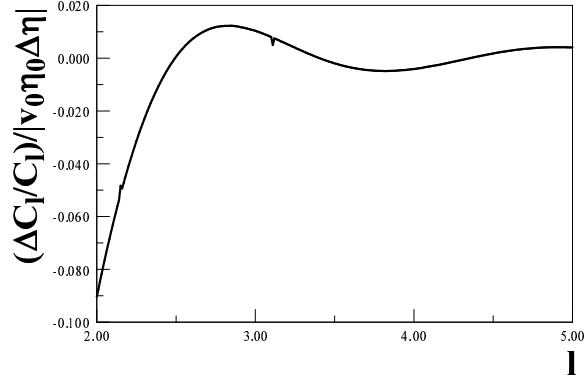}
\caption{ The corrections $(\Delta C_l/C_l)/ |v_0 \; \eta_0\; \Delta| $ vs. $ l $ in the
Born approximation for an attractive potential of the form
(\ref{potential}),   for
$ \kappa \; |\eta_0|=1; \; |\Delta/\eta_0|=0.01 $. The corrections for the
higher multipoles are substantially smaller than the quadrupole correction,
vanishing very rapidly for $l \geq 3$} \label{fig:clborn}
\end{center}
\end{figure}

The corrections for the higher multipoles are substantially smaller,
vanishing very rapidly for $l \geq 3$  as shown in
Fig. \ref{fig:clborn} for the attractive potential.

\subsection{Exactly solvable Potential}\label{sec:well}

The Born approximation is the first order in perturbation theory and
is valid in the regime $ k\gg |v_0| $. A simple and exactly solvable
example is the square well potential
\be
\mathcal{V}(\eta) = \left\{\begin{array}{cc}
                      -|v_0| & \mathrm{for}~~
\eta_1 \leq  \eta \leq  \eta_2  \; , \\
                      0 & \textrm{otherwise}\; , \\
                    \end{array} \right. \; ,
\ee
where $ \eta_1 \equiv \eta_0-\frac{\Delta}2 $ and
$ \eta_2 \equiv \eta_0+\frac{\Delta}2 $,
 $\Delta >0$ is the width of the potential well and $-|v_0|$ its depth.
The case of a potential barrier is obtained by the replacement
$|v_0|\rightarrow -|v_0|$.

We compute the mode functions by matching the functions and
derivatives as in the familiar case of the step-potential in the
one-dimensional Schroedinger equation, and again to lowest order in
slow roll we consider the scale invariant case $ \nu=3/2 $.
The wave function is given by
\be
S(k,\eta) =  \left\{\begin{array}{cc}
g(k\eta)  &  \; , ~~\eta<\eta_1 \\
     \\
E(k)\,g(q\eta) + Q(k)\,g^*(q\eta) &  \; , ~~\eta_1
\leq \eta \leq \eta_2 \; ,\\
   \\
A(k)\,g(k\eta)+B(k)\,g^*(k\eta) &  \; , ~~\eta_2 < \eta <0 \; ,\\
\end{array} \right.
\ee
where
\be
g(k\eta) = e^{-i \, k \, \eta}\Big[1-\frac{i}{k \,
\eta} \Big] ~~;~~ q= \sqrt{k^2+|v_0|}  \; .\label{gs}
\ee
Matching functions and derivatives at  $ \eta = \eta_1 $ and $ \eta = \eta_2 $,
we find
\bea
&&A(k)   =
\frac{e^{i \, k \, \Delta}}{4 \, k \, q}\Bigg\{ e^{-i \, q \,
\Delta} (k+q)^2\Big[1+ i \frac{k-q}{k  \, q \,\eta_1}\Big]\Big[1- i
\frac{k-q}{k \, q\ \,\eta_2}\Big]-  e^{ i \, q \, \Delta}
(k-q)^2\Big[1- i \frac{k+q}{k \, q \, \eta_1}\Big]\Big[1+ i
\frac{k+q}{k \,  q \, \eta_2}\Big] \Bigg\}\label{Asquare}  \; , \cr \cr
&&B(k)   =   - |v_0|\,\frac{e^{-2ik\eta_0} }{4kq}\Bigg\{ e^{-iq\Delta}
 \Big[1+ i \frac{k-q}{k \,  q \, \eta_1}\Big]\Big[1- i \frac{k+q}{k \,
q \, \eta_2}\Big]-  e^{ i \, q \, \Delta}  \Big[1- i \frac{k+q}{k \,
q \, \eta_1}\Big]\Big[1+ i \frac{k-q}{k \,  q \, \eta_2}\Big]
\Bigg\} \label{Bsquare} \; .
\eea
It is straightforward to check the
unitarity condition $ |A(k)|^2 - |B(k)|^2 =1 $. We  consider an
arbitrary depth $|v_0|$ to include non-perturbative aspects, but
focus on the case of a narrow well for which
\be
\Big| \frac{\Delta}{\eta_0}\Big| \ll 1 \label{narrow} \; .
\ee
The number of particles created by the pre-slow roll stage is
\be
N_k = |B(k)|^2 =
\frac{(v_0 \; \eta_0^2)^2}{4(k \; \eta_0)^2}\Big| \frac{\Delta}{\eta_0}
\Big|^2 \Bigg\{\Bigg[\frac{\sin q \, \Delta}{q \; \Delta}
\left(1-\frac{|v_0|}{(k \; q \; \eta_0)^2}\right)-
\frac{\cos q \, \Delta}{(q \; \eta_0)^2} \Bigg]^2+\Bigg[\frac2{k \; \eta_0} \frac{\sin
q \, \Delta}{q \; \Delta} \Bigg]^2 \Bigg\} \label{numbersquare}  \; .
\ee
For arbitrary strength of the potential $v_0$, a small number of
produced particles requires a narrow width (\ref{narrow}).

For $ k \gg |v_0| $ the number of particles is \be \label{Nofksq}
N_k \buildrel{k \to \infty}\over= \frac{v_0^2}{8 \; k^4} \quad . \ee
Moreover, a continuous and differentiable potential $
\mathcal{V}(\eta) $ with smooth edges will yield a $ N_k $ vanishing
faster than $ 1/k^4 $ for large $ k $ since the Fourier transform of
a continuous and differentiable function of compact support falls
off exponentially at large $ k $. Therefore, the asymptotic behavior
for a general continuous potential with a typical scale $ v_0 $ is $
N_k <  v^2_0/k^4 $ and the ultraviolet finiteness of the energy
momentum tensor is guaranteed \cite{motto2}.

\medskip

To leading order in the `narrow width' approximation   the transfer
function is \be D(k) = \frac{|v_0| \,  \eta_0^2 }{k \,
|\eta_0|}\,\Big|\frac{\Delta}{\eta_0} \Big| \Bigg\{ \sin(2 \, k \,
|\eta_0|)\Big[ \frac{\sin q \, \Delta}{q \, \Delta}\left( 1-
\frac{|v_0| \, \eta_0^2}{(k \, q \, \eta_0^2)^2}\right)-\frac{\cos q
\, \Delta}{(q \, \eta_0)^2} \Big] + 2 \; \frac{\cos 2 \, k \,
\eta_0}{k \, |\eta_0|}\,\frac{\sin q \, \Delta}{q \, \Delta}\Bigg\}
\label{Dsquare}  \, , \ee where we have written the transfer
function explicitly in terms of the relevant dimensionless
combinations of parameters $ v_0, \; \Delta $ and $ \eta_0 $.

\bigskip

We have performed an exhaustive numerical study of the corrections
to the $ C_l's $ in a wide range of the dimensionless parameters $
\kappa \; |\eta_0|, \; |v_0| \; \eta_0^2 $ and $ |\Delta/\eta_0| $
where $ \kappa $ is defined in eq.(\ref{kappa}), with the conclusion
that for an attractive potential there is a substantial  suppression
of the quadrupole for $ \kappa \; |\eta_0|\sim 1 $ and that the
corrections  for higher multipoles are negligibly  small and
observationally irrelevant since these fall off as $ 1/l^2 $, hence
much smaller than the irreducible cosmic variance.

\vspace{2mm}

\begin{figure}[h]
\begin{center}
\includegraphics[height=2.5in,width=2.5in,keepaspectratio=true]{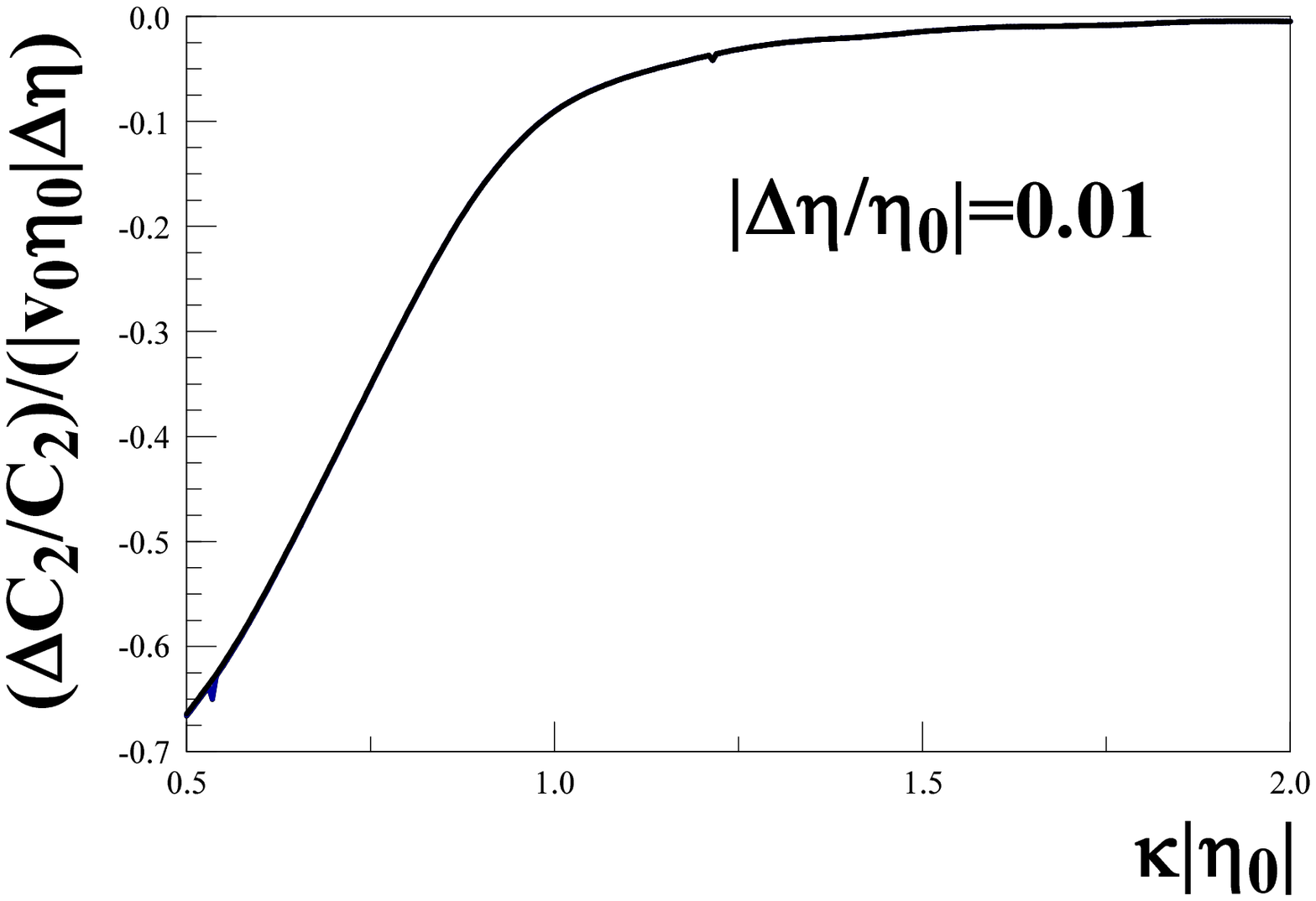}~~~~~~
\includegraphics[height=2.5in,width=2.5in,keepaspectratio=true]{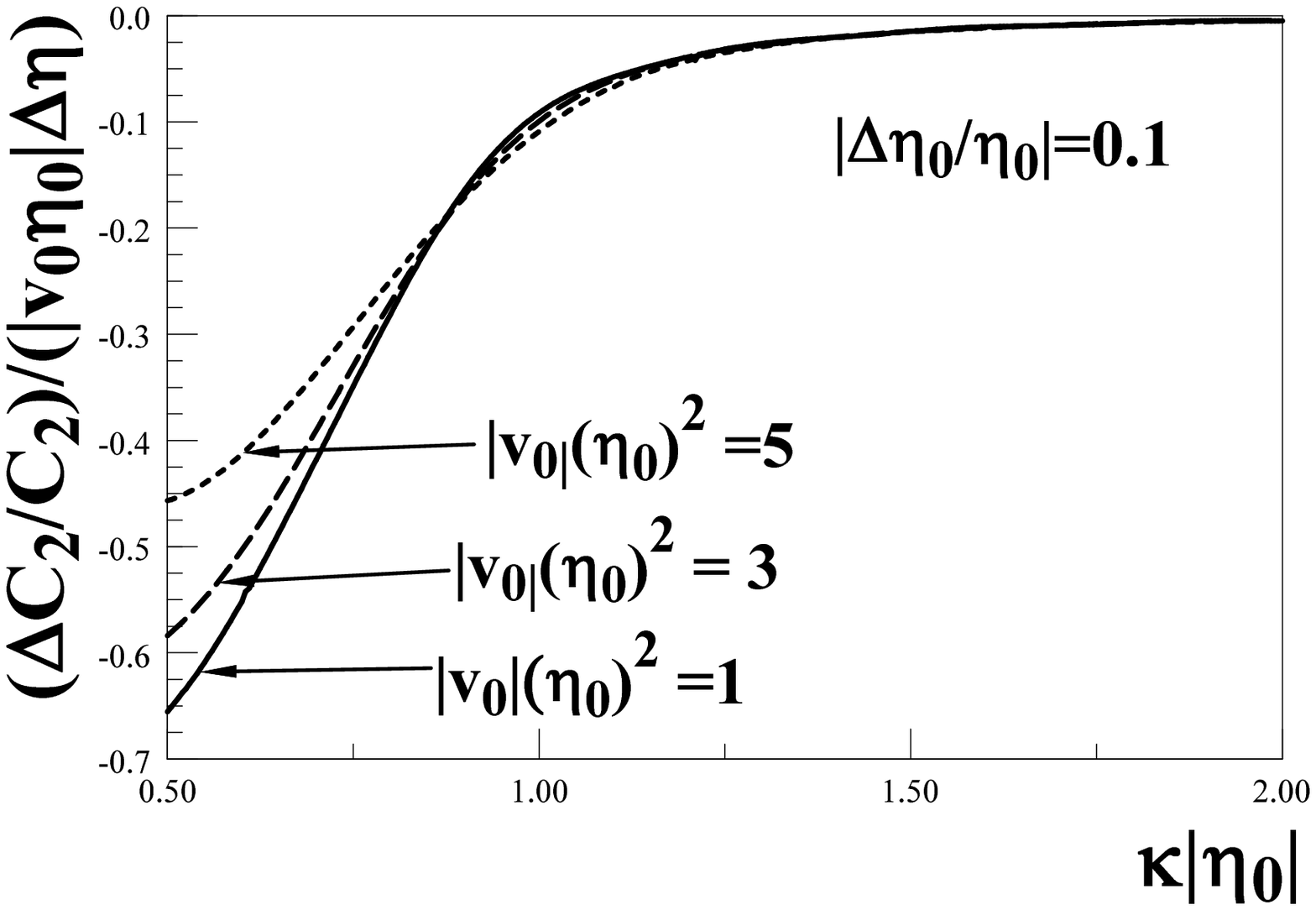}
\caption{ $ (\Delta C_2/C_2)/|v_0 \; \eta_0 \; \Delta| $ vs $
\kappa| \; \eta_0| $  for the square well  potential, for $ |v_0| \;
\eta_0^2 = 1, \; 3, \; 5~;~|\Delta/\eta_0|=0.01 $ (left panel) and $
|\Delta/\eta_0|=0.1 $ (right panel). There is a substantial
suppression of the quadrupole when the potential is localized at a
time scale $ \eta_0 \sim 1/[a_0 \; H_0] $. This time scale is
approximately $55$ e-folds before the end of inflation  when the
wavelengths corresponding to today's Hubble radius exited the
horizon.} \label{fig:c2square}
\end{center}
\end{figure}

Fig. \ref{fig:c2square} displays the changes in the quadrupole for
various representative values of $ |v_0| \; \eta_0^2 $ for the cases
$ |\Delta/\eta_0|=0.01, \; 0.1 $ respectively. It is clear from
these figures that there is a substantial suppression of the
quadrupole if the potential is localized at a time scale $ |\eta_0|
\sim 1/[a_0 \; H_0] $. This time scale approximately coincides  with
$55$ e-folds before the end of inflation when the wavelengths
corresponding to today's Hubble radius exited the horizon.

\vspace{1mm}

Furthermore, for localization scales of the potential $ 0.05 \leq
|\Delta/\eta_0| \leq 0.1 $, a $ 10-20\% $ suppression of the
quadrupole is obtained for $ |v_0| \sim \eta_0^{-2} $. Therefore a
substantial suppression of the quadrupole is explained quite
naturally within the effective field theory of inflation with a
pre-slow roll potential of scale $v_0 \sim \eta_0^{-2}  $.

\begin{figure}[h]
\begin{center}
\includegraphics[height=2.5in,width=2.5in,keepaspectratio=true]{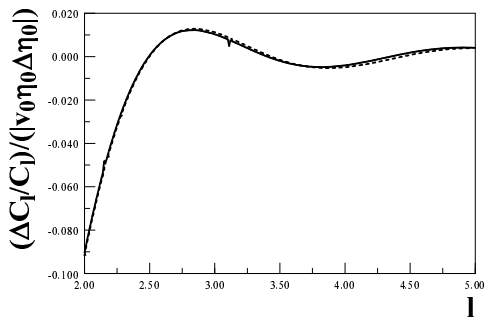}
\caption{ $ (\Delta C_l/C_l)/|v_0 \; \eta_0 \; \Delta| $ vs $ l $  for the
square well  potential  ,   for $ |v_0| \; \eta_0^2 = 1;\kappa \;
|\eta_0|=1 $ and $ |\Delta/\eta_0|=0.01, \; 0.1 $. We see that the
suppresion of higher multipoles are negligibly small (they fall off as $ 1/l^2 $)
and observationally irrelevant.}
\label{fig:clsquare}
\end{center}
\end{figure}

\medskip

These exact results agree remarkably well with the simpler Born
approximation within the range of parameters consistent with small
number of particles as required by the small backreaction condition.
Therefore, this \emph{exactly solvable} example lends support to the
statement that the Born approximation is indeed robust and describes
remarkably well the main corrections from the potential $
\mathcal{V}(\eta) $.

\medskip

These results apply equally well to curvature and tensor perturbations.
Therefore, this analysis leads to the \emph{prediction} that
the quadrupole of \emph{tensor} perturbations will also feature a \emph{suppression}.

\section{Conclusions}

In this article we studied the effect of initial conditions on
metric and tensor perturbations with emphasis on  the observational
consequences of initial conditions   consistent with
renormalizability and small backreaction. Generalized initial
conditions for the mode functions of gauge invariant perturbations
are encoded in Bogoliubov coefficients, or equivalently in
distribution functions of Bunch-Davies quanta. We begun the study by
clarifying the constraint on the Bogoliubov coefficients from the
general restrictions of renormalizability and negligible back
reaction on the energy momentum tensor of gauge
invariant perturbations. These general criteria
constrain the asymptotic behavior for large wave vectors of
the Bogoliubov coefficients up to $ 1/k^4 $. We find that the modifications to the power
spectra of gauge invariant perturbations due to general initial
conditions are encoded in a \emph{transfer function for initial
conditions} $D(k)$.  Our main results are summarized as follows:

\begin{itemize}

\item{ General arguments based on the asymptotic behavior of the Bogoliubov
coefficients at large wave vector show that only the \emph{low}
multipoles, those in the Sachs-Wolfe plateau, are sensitive to
initial conditions allowed by renormalizability and small back
reaction. Effects upon high multipoles are strongly suppressed by
the rapid fall off of the Bogoliubov coefficients at large
wavevectors $k$. We compute the change in the quadrupole due to
generic initial conditions described by the asymptotic limit of the
Bogoliubov coefficients. A substantial change of the order $ 10-20\%
$ on the CMB quadrupole is found when the momentum scale at which
the asymptotic behavior sets corresponds to the physical wavelength
of the order of the Hubble radius \emph{today}. }

\item{ We show that mode functions with general initial conditions
determined before the slow roll stage are equivalent  to those that
result from scattering by the potential $ \mathcal{V}(\eta) $
arising from the cosmological evolution just prior to the onset of
slow roll. The influence of initial conditions upon the power
spectra of curvature and tensor perturbations is encoded in a
transfer function $D(k)$. }

\item{Implementing methods
from scattering theory, we develop the formulation of initial
conditions arising from  scattering by a potential $ \mathcal{V}(\eta) $
and obtain the transfer function of initial conditions
 $ D(k)$ in terms of this potential.
The changes in the low multipoles are studied both in the Born
approximation and in a exact solvable case for $\mathcal{V}(\eta)$,
with  complete  agreement between the results in both cases. The
transfer function for initial conditions is shown to have the
asymptotic large $k$ behavior consistent with renormalizability and
negligible back reaction.}

\item{Furthermore, this
study reveals that \emph{attractive} potentials lead to a
suppression of the quadrupole with a value consistent with the WMAP
data if the the potential is localized at a time scale $\eta_0$ very
near the scale at which the wavelength corresponding to the Hubble
radius \emph{today} exits the horizon during inflation, with a
strength $\mathcal{V}(\eta_0) \sim 1/\eta^2_0$. The change in the
$l$-multipole falls off as $ 1/l^2 $ as a consequence of the fall
off of the Bogoliubov coefficients for large $ k $. This entails
that only the quadrupole features an observable suppression, while
the corrections in higher multipoles are not observable with the
present data.}

\item{Our study applies   to   curvature and tensor perturbations, hence
we predict a suppression of  quadrupole for B-modes for an
attractive   potential localized prior to slow roll.}

\end{itemize}

In the companion article \cite{II} we show   that the potential
$\mathcal{V}(\eta)$ which determines the initial conditions for the
fluctuations in the slow roll stage is a general feature of a stage
of \emph{fast roll}  inflaton dynamics  followed by slow roll. Under
general circumstances this potential   turns out to be attractive
and results in  a {\bf suppression} on the CMB quadrupole of the
order $ \sim 10-20 \% $ consistent with the observations.

\begin{acknowledgments} D.B.\ thanks the US NSF for support under
grant PHY-0242134,  and the Observatoire de Paris and LERMA for
hospitality during this work. He also thanks Rich Holman for an
initial conversation on initial conditions, and John P. Ralston for
illuminating conversations and challenging prodding.
\end{acknowledgments}

\end{document}